\documentclass[a4paper,fleqn]{cas-dc}

\usepackage[authoryear,longnamesfirst]{natbib}

\def\tsc#1{\csdef{#1}{\textsc{\lowercase{#1}}\xspace}}
\tsc{WGM}
\tsc{QE}

\usepackage{mathtools,ulem,hyperref}
\DeclareMathOperator*{\concat}{\Vert}
\newcommand{\norm}[1]{\Vert#1\Vert}
\DeclareMathOperator*{\mean}{mean}
\newcommand{\revision}[2]{#2}
\newcommand{\minorrevision}[2]{#2}

\begin{document}
\let\WriteBookmarks\relax
\def\floatpagepagefraction{1}
\def\textpagefraction{.001}

\shorttitle{Deep vectorised operators for hemodynamics estimation}    

\shortauthors{Suk et al.}  

\title [mode = title]{Deep vectorised operators for pulsatile hemodynamics estimation in coronary arteries from a steady-state prior}  



%

\author[1]{Julian Suk}[orcid=0000-0003-0729-047X]

\cormark[1]


\ead{j.m.suk@utwente.nl}



\affiliation[1]{organization={Department of Applied Mathematics and Technical Medical Center, University of Twente},
            city={Enschede},
            country={The Netherlands}}

\author[2]{Guido Nannini}





\affiliation[2]{organization={Department of Electronics Information and Bioengineering, Politecnico di Milano},
            city={Milan},
            country={Italy}}

\author[1]{Patryk Rygiel}





\author[1]{Christoph Brune}





\author[3,4]{Gianluca Pontone}





\affiliation[3]{organization={Department of Perioperative Cardiology and Cardiovascular Imaging, Centro Cardiologico Monzino IRCCS},
	city={Milan},
	country={Italy}}
\affiliation[4]{organization={Department of Biomedical, Surgical and Dental Sciences, University of Milan},
	city={Milan},
	country={Italy}}

\author[2]{Alberto Redaelli}





\author[1]{Jelmer M. Wolterink}





\cortext[1]{Corresponding author}



\begin{abstract}
\textit{Background and objective:} Cardiovascular hemodynamic fields provide valuable medical decision markers for coronary artery disease. Computational fluid dynamics (CFD) is the gold standard for accurate, non-invasive evaluation of these quantities \textit{in \revision{vivo}{silico}}. In this work, we propose a time-efficient surrogate model, powered by machine learning, for the estimation of pulsatile hemodynamics based on steady-state priors.\\
\textit{Methods:} We introduce deep vectorised operators, a modelling framework for discretisation-independent learning on infinite-dimensional function spaces. The underlying neural architecture is a neural field conditioned on hemodynamic boundary conditions. Importantly, we show how relaxing the requirement of point-wise action to permutation-equivariance leads to a family of models that can be parametrised by message passing and self-attention layers. We evaluate our approach on a dataset of 74 stenotic coronary arteries extracted from coronary computed tomography angiography (CCTA) with patient-specific pulsatile CFD simulations as ground truth.\\
\textit{Results:} We show that our model produces accurate estimates of the pulsatile velocity and pressure (approximation disparity 0.368 $\pm$ 0.079) while being agnostic ($p < 0.05$ in a one-way ANOVA test) to re-sampling of the source domain, i.e. discretisation-independent.\\
\textit{Conclusions:} This shows that deep vectorised operators are a powerful modelling tool for cardiovascular hemodynamics estimation in coronary arteries and beyond.
\end{abstract}





\maketitle

\section{Introduction}
\the\textwidth
\the\columnwidth
Coronary artery disease (CAD) consists in the accumulation of plaque in the artery wall which results in narrowing of the blood vessel. Stenoses then impair
the flow of oxygenated blood through the coronary arteries to the heart muscle, which can lead to myocardial infarction~\citep{ZingaroVergara2023}. Visualisation and quantification of patient-specific blood flow can be valuable for diagnosis~\citep{DriessenDanad2019}, prognosis~\citep{CandrevaPagnoni2022} and treatment~\citep{ChungCebral2015} of cardiovascular disease. Quantities derived from the blood flow and pressure, e.g., wall shear stress, oscillatory shear index and fractional flow reserve (trans-stenotic pressure ratio, measured during maximal hyperemia),
have been identified as useful decision markers and risk factors in CAD. \revision{}{For instance, wall shear stress can act as a predictor for culprit lesions in cases in which fractional flow reserve is above threshold (> 0.8) but the lesion is ischemic~\citep{LeeChoi2019}. Wall shear stress and other biomarkers derived from blood flow are indicated as \textit{possible alternatives} to detect ischemic lesions.} Computing these biomarkers requires access to pulsatile velocity and pressure fields within the vessel.

Clinical hemodynamic biomarkers can be non-invasively quantified \textit{in-vivo} with imaging techniques such as Doppler ultrasound, particle image velocimetry or 4D flow magnetic resonance imaging (MRI). However, the small size of the coronary arteries limits the extraction of hemodynamic metrics that can effectively distinguish pathological conditions. Alternatively, these metrics can be obtained \textit{in-silico} via computational fluid dynamics (CFD) based on 3D patient models reconstructed from computed tomography (CT)~\citep{Nannini2024}. Some of the challenges related to clinical adoption of CFD are its long runtimes, high computational demand and dependence on expert knowledge throughout the process.  
Furthermore, sensitivity to modelling choices, such as \textit{discretisation} of time and space, as well as boundary conditions, make it difficult to decide upon systematic protocols that are consistent and repeatable across hospitals. Indeed, it has been shown that there is inter-operator variability between CFD simulations based on identical medical images~\citep{ValenSendstadBergersen2018}.
While CFD is gradually making its way into the clinic, pioneered in part by commercial service providers like HeartFlow~\citep{WuWu2024},
the above challenges make such services expensive which may hinder widespread adoption.

In this work, we address these challenges with an in-silico surrogate model for estimation of coronary hemodynamics, based on machine learning. We train the model on a dataset of pulsatile CFD simulations, i.e. time-varying velocity and pressure fields, in right and left coronary branches of 74 patients scheduled for clinically indicated invasive coronary angiography for suspected CAD. Additionally, we create a steady-state simulation for each patient, representing the pulsatile CFD solution at the first instant of the cardiac cycle (i.e., at time $t = t_0 + \mathrm{d}t$ where $t_0$ is the end diastolic time) and use them to cast the learning objective as operator learning: we let our neural networks learn a mapping from the steady-state to the corresponding pulsatile hemodynamics. The steady-state CFD solution provides a powerful prior that enables us to faithfully predict the pulsatile hemodynamics even with limited (yet diverse) data.

We name our approach \textit{deep vectorised operators} and lay out a modelling framework in which the nonlinear operator is learned by a vectorised conditional neural field. Relaxing the requirement of point-wise action of neural fields to permutation equivariance allows us to parametrise our models as PointNet++~\citep{QiYi2017} and transformers~\citep{VaswaniShazeer2017} which have recently proven capable of learning hemodynamics in large-scale, anatomical meshes~\citep{MoralesFerezMill2021,SukHaan2024b}. To further facilitate learning from limited amounts of data, we incorporate group equivariance in our models, which can be advantageous for extracting information from small hemodynamic datasets~\citep{SukBrune2023}. As in CFD~\citep{LopesPuga2020}, our method is informed by patient-specific boundary conditions.
Once trained, our models can quickly produce pressure and velocity field estimates for vascular anatomy on consumer hardware based on boundary conditions for a new patient. We empirically show that our model is robust w.r.t. discretisation of the flow domain.

Our contributions are as follows: (1) we explore learning pulsatile hemodynamics from steady-state priors with a nonlinear operator, (2) we provide a unified modelling framework for neural hemodynamic field estimation with state-of-the-art neural architectures and (3) we evaluate our approach on a relevant, real-life dataset of coronary arteries with patient-specific CFD simulations.

\subsection{Related works}
Estimation of hemodynamics with machine learning methods has been an active area of research. Typically, these works fall into one of two categories: (1) transductive instance optimisation with physics-based regularisation (solving partial differential equations) and (2) inductive, generalising feed-forward methods that learn to infer hemodynamics based on patient anatomy and boundary conditions. The former category requires long training times for each new subject, often in the same order of magnitude as CFD. These methods can be useful in scenarios where hemodynamic parameters are only partially known~\citep{RaissiYazdani2020,FathiPerezRaya2020,KontogiannisJuniper2021}. In contrast, methods in the latter category are designed as fast, compute-efficient CFD surrogates.

In this study, we concentrate on the second approach: leveraging neural networks to model the relationship between vascular geometry, boundary conditions and hemodynamic patterns. These neural networks process 3D vessel structures to predict blood-flow-related parameters on the vessel surface or within the interior. In addition to research aimed at predicting 1D quantities along the vessel tree~\citep{PegolottiPfaller2024} and surface characteristics such as wall shear stress~\citep{SukHaan2024a,GharleghiSowmya2022}, pressure drop~\citep{RygielPluszka2023} as well as endothelial cell activation potential~\citep{MoralesFerezMill2021}, several studies have focused on predicting volumetric vector fields. For example, \cite{LiangMao2020} and \cite{WangWu2023} trained fully-connected neural networks that work on 3D point-cloud representations of the carotid artery and thoracic aorta, respectively, to predict pressure and velocity fields. \cite{MaulZinn2023} developed an octree-based neural network, paired with trilinear interpolation to learn parametric, pulsatile flow independently of spatial discretisation and applied it to synthetic vascular trees. \cite{LiWang2021} employed a point-cloud-based architecture to estimate pressure and velocity fields in coronary arteries and synthetic cerebral aneurysms. \cite{ZhangMao2023} used a combination of PointNet++~\cite{QiYi2017} and a physics-informed neural network (PINN) based on the Navier-Stokes equations to estimate velocity fields in 3D models of the abdominal aorta. In prior work~\citep{SukBrune2023} we used a multiscale, E(3)-steerable graph neural network to estimate velocity fields in synthetic coronary arteries.

\revision{}{\cite{WesselsKnigge2025} recently proposed equivariant neural fields by cross-conditioning on latent geometric attributes using transformers. This differs from our approach in that the field domain and geometric object are decoupled while in our case they coincide.}

Operator learning approaches, such as DeepONet~\citep{LuJin2021} and (Fourier) neural operators~\citep{KovachkiLi2024,LiKovachki2021} have recently gained traction in the machine learning community but there are few applications to cardiovascular hemodynamics estimation. \revision{}{\cite{AlkinFürst2024} recently proposed a transformer-based encoder-decoder model to learn neural operators for general physical systems.}

\begin{figure*}
	\centering
	\includegraphics{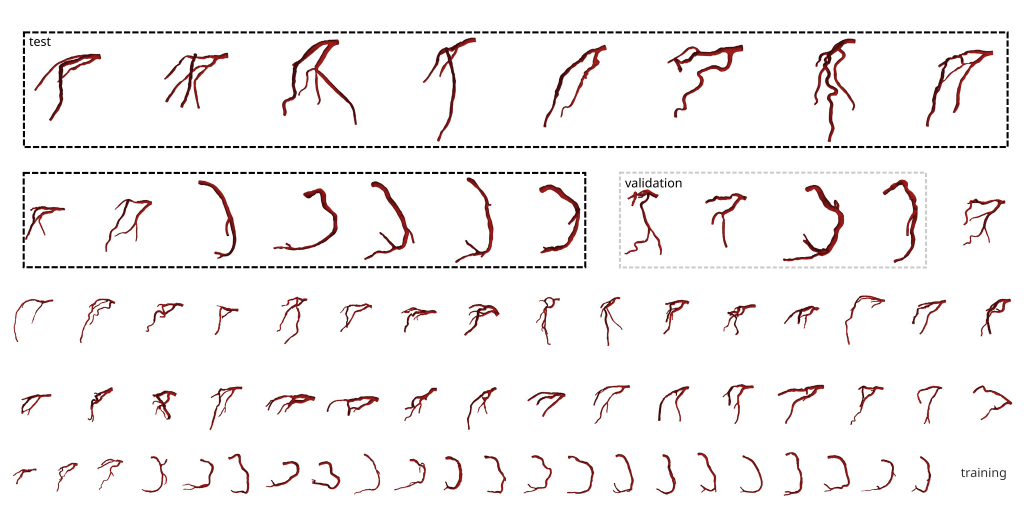}
	\caption{\revision{}{\textbf{Coronary artery dataset} consisting of 74 3D models of stenotic left and right coronary arteries with different numbers of bifurcations and outlets. Patients in this cohort were scheduled for invasive coronary angiography due to suspected CAD. We partitioned the dataset into training, validation and test split while ensuring equal ratios of left and right coronary arteries between training and test split.}}\label{fig:mosaic}
\end{figure*}

\section{Materials and methods}
\subsection{Coronary artery dataset}
Patient-specific CFD simulations of coronary flow from a previous study~\citep{Nannini2024} served as the dataset. A brief overview of the data generation process is provided in the following. Coronary computed tomography angiography (CCTA) scans of patients scheduled for clinically-indicated invasive coronary angiography due to suspected CAD were retrospectively collected from Centro Cardiologico Monzino (Milan, Italy). Overall, 74 stenotic vessel geometries \revision{}{(shown in Figure~\ref{fig:mosaic})}, from both the left (48) and right (26) coronary arteries, were reconstructed from CCTA scans using 3DSlicer~\citep{Fedorov2012}. The segmentation included the artery branches downstream of the ostium, and was terminated when the vessel diameter fell below 1.5 mm, in accordance with established guidelines~\citep{Pontone2017}. The aortic root was excluded. The study was performed in accordance with recommendations of the local Ethics Committee, with written informed consent from all subjects, in accordance with the Declaration of Helsinki.

The volume of the reconstructed geometries was discretized \revision{}{-- after mesh sensitivity analysis (details in supplementary material) --} into tetrahedral elements with a characteristic size of 0.25 mm using the TetGen algorithm embedded within SimVascular~\citep{SimVascular2017}. This led to 1,801,029 tetrahedral elements and 314,525 vertices on average. For each mesh, three cardiac cycles were simulated by numerically solving the incompressible Navier-Stokes equations, with time-varying boundary conditions, in SimVascular \revision{}{using \texttt{svSolver} while neglecting any effect related to arterial wall motion and deformation}. A time discretization of $\mathrm{d}t=0.001$ seconds was adopted in the CFD, results were exported every 0.025 seconds, which corresponds to the largest time step value that ensured the numerical convergence of the simulation (based on separately run sample-tests). Results were retained solely from the last cycle, spanning 35 time steps on average.

To set up the simulation, the average hyperemic flow rate, $\bar{Q}_\text{hyp}$, was computed \revision{from baseline clinical measurements and}{using clinical data. Briefly, the myocardial mass $M_\text{myo}$, heart rate $H$ and systolic pressure ($P_\textit{sys}$) were used to compute the resting flow rate using a formula by \cite{SharmaItu2012}: $Q_\textit{rest} = 0.14 \cdot (7 E - 4 \cdot H \cdot P_\textit{sys} - 0.4) \cdot M_\text{myo}$. The hyperemic flow was set to 3.5-fold the resting one and} used to scale a flow rate waveform obtained from the literature~\citep{Flemister2020}, which was applied as the inlet boundary condition (BC) $Q_\text{hyp}(t)$ in the simulation. A five-element Windkessel model was coupled to each outlet as the BC, with its parameters tuned \revision{based on the patient's aortic pressure}{to match the patient’s mean aortic pressure, for resistance, and pulse pressure, for compliance} (for further details on the CFD settings, see \citep{Nannini2024}). Pulsatile simulations took between 12 to 24 h per case \revision{}{on a 40 cores Intel Xeon CPU X5670 machine with 64.4 GB RAM}.

Additionally, a steady-state CFD simulation was performed for each model using the inflow value corresponding to the flowrate at the first pulsatile-CFD solution output: $Q_\text{hyp}(t=0.025)$.
We chose this initial time point for the steady-state simulation to minimize inertial effects, which can influence the pressure field in pulsatile CFD simulations. Simulations took between 5 to 30 min depending on the flow rate where computational cost scaled roughly with the Reynolds number. \revision{}{In the context of our machine learning model, the steady-state solution functions as cheap information about the characteristic flow in the respective artery. It can be thought of as the fluid-dynamic system response to an ``impulse excitation''.}

\subsection{Background}
In the following, we will discuss the relevant theoretical background to put our approach into context. However, our method can be understood independently of this section.

In this work we consider the problem of learning time-dependent, $c$-dimensional vector fields ($c \in \mathbb{N}$) from data
\[
y \colon \begin{cases}
	T \times \Omega \to \mathbb{R}^c\\
	t, x \mapsto y(t, x)
\end{cases} T \subset \mathbb{R},\hspace{12pt} \Omega \subset \mathbb{R}^3,
\]
which are defined on a time interval $T$ and spatial domain $\Omega$.
In the following we propose a family of models that can efficiently represent specific instances of such vector fields while enabling zero-shot \textit{generalisation} to instances beyond the training data. In contrast to common applications in medical imaging, we do not assume the data to have grid structure that we can exploit, e.g., via convolutional neural networks (CNN). Rather, our model learns a map between infinite-dimensional function spaces.

\subsubsection{Neural fields}
A neural field is a parametrised map $f_\theta \colon \mathbb{R}^{c_\text{in}} \to \mathbb{R}^{c_\text{out}}$ between
vector spaces where $f_\theta$ is a neural network with parameters $\theta$, e.g., a multilayer perceptron (MLP). The domain $\mathbb{R}^{c_\text{in}}$ is commonly required to be spacetime~\citep{XieTakikawa2022}. The neural network $f_\theta$ is trained under observations $\{(x^i \in \mathbb{R}^{c_\text{in}}, y^i \in \mathbb{R}^{c_\text{out}})\}_{i \in [1, n]}$ of input-output pairs.\footnote{\revision{}{$c_\text{in}$ and $c_\text{out}$ are the dimensions of domain and co-domain (respectively) and $n$ is the total number of observations.}} Since neural networks are commonly continuous or continuously differentiable, neural fields can be smooth representations of scalar and vector fields in a variety of applications, such as neural signed distance functions~\citep{GroppYariv2020} and physics-informed neural networks~\citep{RaissiPerdikaris2019}.

\paragraph{Conditional neural fields}
Neural fields can be conditioned on a collection of parameters $\xi \in \Xi$, e.g., by concatenating ($\concat$) them with the input observations and feeding them to the neural network
\[
f_\theta \colon \begin{cases}
	\mathbb{R}^{c_\text{in}} \times \Xi \to \mathbb{R}^{c_\text{out}}\\
	x, \xi \mapsto f_\theta(x \concat \xi)
\end{cases}.
\]
This enables, e.g., representing multiple vector fields with the same domain and co-domain in a single neural field by assigning each vector field a latent code, as proposed by DeepSDF~\citep{ParkFlorence2019}, or representing \textit{parametric} solutions to partial differential equations. In this case, observations depend on the parameters $\xi$ under which they were made.

\paragraph{Vectorised conditional neural fields}
When training conditional neural fields, input observations $x^i$ are commonly vectorised into an ($n \times c_\text{in}$)-dimensional tensor while $f_\theta$ (conditioned on $\xi$) broadcasts over the batch dimension.  
Accordingly, $f_\theta$ must act point-wise, i.e., row-wise on the input tensor.
However, if we always feed observations to the neural field as complete sets, we can relax this requirement and choose $f_\theta$ -- instead of point-wise -- as point-wise \textit{permutation-equivariant}. Such models allow introducing inter-observation context and thus implicit conditioning on the complete set of observations. Based on the same idea, \cite{HagnbergerKalimuthu2024} proposed this family of models under the name ``vectorised conditional neural fields''. \revision{}{This inter-observation context conditions the neural field on characteristics of the input (e.g. topology or global statistics) and supports interactions in input spacetime. Precisely this cross-conditioning gives rise to the expressive power of set-based neural networks like PointNet++ and transformers. As drawback of adding context, sets must always be processed in their entirety. In contrast, point-wise independent processing natively enables sub-dividing the sets into chunks which is useful in scenarios where memory is a bottleneck. We identify permutation equivariance as the key requirement with which set-based neural networks can be cast as \textit{vectorised} neural fields.}

\subsubsection{Permutation-equivariant neural networks}
Permutation-equivariant functions satisfy the condition
\[
(f_\theta \circ P)(\cdot) \equiv (P \circ f_\theta)(\cdot)\hspace{12pt} \text{for all}\hspace{12pt} P \in \mathrm{S}_n,
\]
where $\circ$ denotes composition and $\mathrm{S}_n$ is the symmetric group of permutations on $n$ set elements. In other words, the relation between input and output elements cannot depend on their specific ordering. Note that point-wise functions are trivially permutation-equivariant. Among the most popular permutation-equivariant building blocks for neural network composition are message passing and self-attention.  

\paragraph{Message passing}
In this work we consider message passing on point clouds. Message passing layers act on local neighbourhoods $\mathcal{N}_p$ around points $p$ in the input vector space by constructing messages and updating point features $h^p$
\begin{align*}
	m_{p, q} &= \mathrm{MLP}(h^p \concat h^q \concat p \concat q), &\text{(message from $q$ to $p$)}\\
	h^p &\leftarrow \bigoplus\limits_{q \in \mathcal{N}_p} m_{p, q} &\text{(point feature update)}
\end{align*}
where $\bigoplus$ is an aggregation operator like mean or maximum \revision{}{and $\concat$ denotes concatenation}. Permutation equivariance is enabled by the dependence of the local neighbourhoods $\mathcal{N}_p$ on the input vector space.

\begin{figure*}
	\centering
	\includegraphics{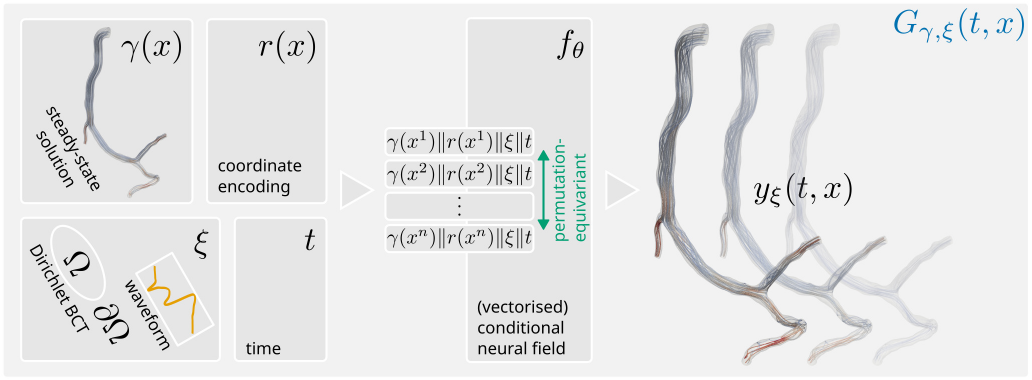}
	\caption{\textbf{\revision{}{Deep vectorised operator.}} Steady-state hemodynamic fields $\gamma(x)$ and relational point encoding $r(x)$ are fed to a permutation-equivariant model $f_\theta$ together with conditions $\xi$ and query time $t$. Functional quantities $\gamma(x)$ and $r(x)$ can be vectorised, i.e., supplied all at once, to enable cross-conditioning by parametrising $f_\theta$ with message passing or self-attention layers. In that case, global quantities $\xi$ and $t$ are broadcast over the batch dimension. Neural field $f_\theta$ then predicts pulsatile hemodynamics $y_\xi(t, x)$ at (each of) the query point(s) $(t, x)$ in spacetime. This defines the nonlinear operator $G_{\gamma, \xi}(t, x)$.}\label{fig:zero}
\end{figure*}

\paragraph{Self-attention}
Given an ($n \times c_\text{in}$)-dimensional tensor $x$, multi-head self-attention layers can be defined as follows:
\begin{align*}
	a_i &= \mathrm{Softmax}\left( \frac{q_i(x) k_i(x)^\mathsf{T}}{\sqrt{c_\text{in}}} \right) v_i(x)\\
	A &= x + \mathrm{Linear}\left(\concat_i a_i\right)\\
	x &\leftarrow A + \mathrm{MLP}(A),
\end{align*}
where Softmax acts entry-wise and we require query, key and value functions $q_i, k_i, v_i$ as well as Linear and MLP to be row-wise permutation-equivariant. \revision{}{We denote by $\Vert_i$ sequential concatenation.} Self-attention is a composition of permutation-equivariant operations and computes pair-wise interactions between all $n$ set elements.

\subsubsection{$\mathrm{E}(3)$-equivariant neural networks}
Apart from being equivariant under permutations of set elements, neural networks can be equivariant under element-wise rotation, translation and reflection in the input vector space $\mathbb{R}^3$ if they satisfy the condition
\[
(f_\theta \circ \rho)(\cdot) \equiv (\rho \circ f_\theta)(\cdot)\hspace{12pt} \text{for all}\hspace{12pt} \rho \in \mathrm{E}(3).
\]
In other words, the relation between input and output elements cannot depend on their specific location and orientation in ambient space. All possible rotations, translations and reflections constitute the Euclidean group E(3).

\subsubsection{DeepONet and MIONet}
DeepONet~\citep{LuJin2021} and its extension MIONet~\citep{JinMeng2022} are frameworks for learning operators between function spaces. Let $\gamma$ be a source and $y$ be a target function. The objective is to learn an operator $G_\gamma \colon \mathbb{R}^{c_\text{in}} \to \mathbb{R}^{c_\text{out}}$ so that $G_\gamma(x) \equiv y(x)$. MIONet generalises this to multiple source functions which may include boundary conditions $\xi$, i.e.,
\[
G_{\gamma, \xi}(x) \equiv y_\xi(x).
\]
Learning this operator is realised by feeding observations of the input functions to ``branch'' networks $b_1, b_2$ and coordinates $x$ to a ``trunk'' network $\tau$ and multiplying their outputs entry-wise:
\[
G_{\gamma, \xi}(x) = \mathrm{Sum}\left(b_1\left(\{\gamma^i\}_{i \in [1, n]}\right) \odot b_2\left(\{\xi^i\}_{i \in [1, l]}\right) \odot \tau(x)\right)
\]
where Sum denotes summations of all vector entries and $\odot$ is the Hadamard product. Optionally, a bias term can be added.

\subsection{Deep vectorised operator}
In this work, we propose a model that takes as inputs steady-state hemodynamic fields as well as pulsatile boundary conditions and estimates the pulsatile hemodynamic fields. Let $\Omega$ be the artery and denote by $u, p$ velocity and pressure, respectively. Define the steady-state hemodynamic fields as
\[
\gamma \colon \begin{cases}
	\Omega \to \mathbb{R}^3 \times \mathbb{R}\\
	x \mapsto u \concat p
\end{cases}
\]
and the pulsatile hemodynamic fields as
\[
y_\xi \colon \begin{cases}
	T \times \Omega \to \mathbb{R}^3 \times \mathbb{R}\\
	t, x \mapsto u_\xi \concat p_\xi
\end{cases}, \xi \in \Xi
\]
where $\Xi$ is the function space of all pulsatile boundary conditions described by three quantities: velocity, pressure at the artery inlet and pulsatile waveform.
We allow $\xi$ to be dependent on $t$ which we omit for compactness of notation. Conceptually, our model is in between MIONet and (vectorised) conditional neural fields. Our objective is to learn an operator
\[
G_{\gamma, \xi} \colon T \times \Omega \to \mathbb{R}^3 \times \mathbb{R}\hspace{12pt} \text{so that}\hspace{12pt} G_{\gamma, \xi}(x) \equiv y_\xi(x).
\]
We do so via a conditional neural field $f_\theta$ so that point-wise
\[
G_{\gamma, \xi}(t, x) = f_\theta(\gamma(x) \concat t \concat r(x) \concat \xi)
\]
where $r(x)$ is fixed, relational encoding of coordinates which we describe below. We train $f_\theta$ on tuples of observations
\[
\{(\gamma(x^j), t^i, x^j, y_\xi(t^i, x^j)) \mid i \in [1, m], j \in [1, n]\}
\]
in a vectorised manner. Thus, we can introduce spatial context by allowing $f_\theta$ to be permutation-equivariant. This setup can also be viewed as a time-continuous, space-discrete, conditional neural field.

\begin{table*}
	\caption{\textbf{Quantitative evaluation of backends.} We report mean $\pm$ standard deviation (across time and subjects) of Approx. disp. and Cos. similarity w.r.t. hemodynamics $u \concat p$ jointly, as well as velocity $u$ and pressure $p$ separately. We denote in \textbf{bold} best evaluation metrics where they significantly ($p < 0.05$) differ from the next best (one-way ANOVA test). Additionally, we state each model's training complexity.}\label{tab:backend}
	\centering
	\renewcommand{\arraystretch}{1.3}
	\begin{tabular}{@{}llcccc@{}}
		\toprule
		&& $u \concat p$ & \multicolumn{2}{c}{$u$} & $p$\\
		\cmidrule(lr){3-3} \cmidrule(lr){4-5} \cmidrule(lr){6-6}
		Backend & Train. complex. & Approx. disp. $\downarrow$ & Approx. disp. $\downarrow$ & Cos. similarity $\uparrow$ & Approx. disp. $\downarrow$\\
		\midrule
		MLP & $254$ s/epoch$\hphantom{^*}$ & $0.432 \pm 0.065$ & $0.439 \pm 0.066$ & $0.72 \pm 0.05$ & $0.121 \pm 0.068$\\
		PointNet++ & $164$ s/epoch$\hphantom{^*}$  & $0.382 \pm 0.073$ & $0.388 \pm 0.073$ & $\mathbf{0.76} \pm 0.04$ & $0.115 \pm 0.075$\\
		LaB-GATr & $857$ s/epoch$^*$  & $\mathbf{0.368} \pm 0.079$ & $\mathbf{0.374} \pm 0.080$ & $\mathbf{0.76} \pm 0.04$ & $\mathbf{0.105} \pm 0.049$\\
		LaB-VaTr & $191$ s/epoch$\hphantom{^*}$  & $0.402 \pm 0.076$ & $0.409 \pm 0.076$ & $0.74 \pm 0.04$ & $\mathbf{0.098} \pm 0.072$\\
		\bottomrule
		\multicolumn{6}{l}{$^*$on four devices in parallel}
	\end{tabular}
\end{table*}

\subsubsection{Backend variants}
Conditional (vectorised) neural field $f_\theta$ can be embodied by \revision{a plethora of}{many different} neural architectures. When designing neural networks for large-scale hemodynamic problems, it is imperative that the models are scalable in the number of query points w.r.t. computation complexity and device usage. We choose a simple, point-wise MLP, the permutation-equivariant PointNet++ and the permutation- as well as SE(3)-equivariant LaB-GATr~\citep{SukImre2024,SukHaan2024b} with memory-efficient attention~\citep{RabeStaats}.
We also include LaB-VaTr, which uses the same layout as LaB-GATr but \textbf{va}nilla fully-connected layers instead of geometric algebra.

\subsubsection{Coordinate encoding}
We equip our model with a powerful coordinate space by lifting $x^j$ to a higher-dimensional, relational encoding $r(x^j)$ of spatial location.\footnote{\revision{}{Despite its similar name, the purpose of this coordinate encoding is to construct canonical coordinates across arteries, different from positional encoding for combatting spectral bias~\citep{RahamanBaratin2019}.}} To this end, we compute -- for every point $x^j$ within the artery --  relative position to the inlet, closest outlet and inner lumen wall~\citep{SukBrune2023}. We separate these vectors into magnitude and direction, both of which we feed to $f_\theta$. Furthermore, we employ a geometry-aware location descriptor in the form of \textit{diffusion distance} and direction to inlet and outlets. For this we use the heat method for distance computation~\citep{CraneWeischedel2017}, originally developed for geodesic distance on surfaces, which readily transfers to volumetric point clouds. The heat method works with a point cloud Laplacian to define an analogue to the heat equation. We can then \revision{``simulate''}{simulate} diffusion of heat sources placed at the inlet and outlets, respectively, and treat the resulting heat map as distance field. \revision{}{Solving the heat equation on point clouds requires solution of a sparse linear system which can be done in near-linear time~\citep{CraneWeischedel2017}.} To obtain the ``direction'' of  diffusion, we simply take the gradient of said field via the discretisation scheme described in \cite{SukAlblas2024}.

\subsubsection{Boundary condition encoding}
As conditions $\xi$, we supply to our model at each point in time $t$ the average velocity vector over the inlet, decomposed into magnitude and direction, along with the average pressure over the inlet. We assume access to both these quantities because they can readily be measured in hospitals. Additionally, we assume access to a patient-specific blood flow waveform of which we query the mean, standard deviation, minimum and maximum value over the cardiac cycle, which we feed to our model along with the subject's heart rate. \revision{}{We do so by simple concatenation to the point-wise observations of the functional quantities $\gamma(x)$ and $r(x)$ (compare Figure~\ref{fig:zero}).}

\subsubsection{Pressure drop estimation}
Instead of estimating point-wise pressure $p(t, x)$ directly, we found it beneficial to predict the \textit{pressure drop} relative to the pressure at the inlet.

\begin{figure*}
	\centering
	\includegraphics{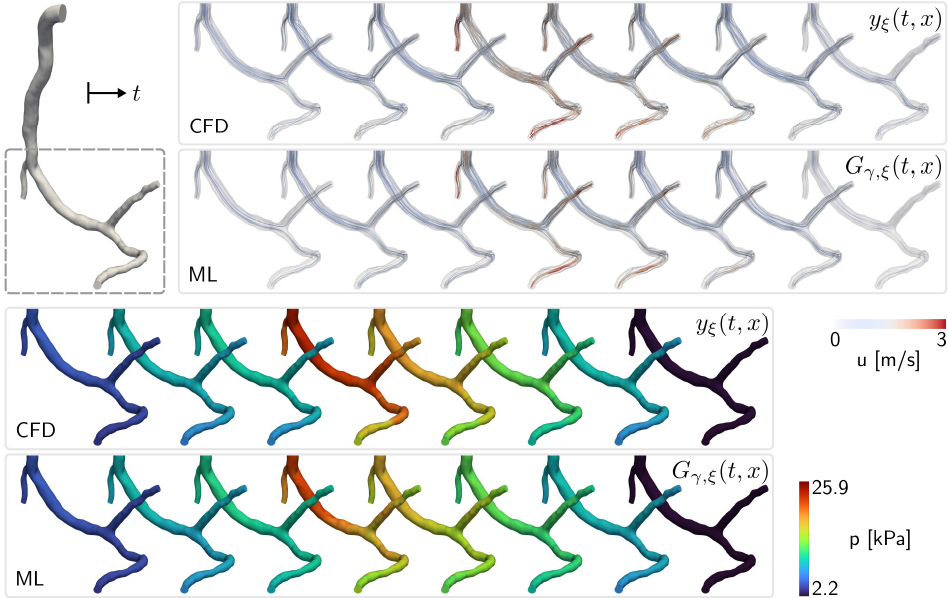}
	\caption{\textbf{Qualitative evaluation} of machine learning (ML) model $G_{\gamma, \xi}(t, x)$ in comparison to CFD $y_\xi(t, x)$ in a test subject (generalisation). Approx. disp. associated with $u \concat p$ estimation in this subject was $0.325$.  
		We visualise streamlines and pressure maps for eight time points in the cardiac cycle with spacing $\mathrm{d}t = 0.125$ s. The backend model was LaB-GATr.}\label{fig:qualitative}
\end{figure*}

\subsection{Evaluation metrics}
To evaluate our method, we must compare vector fields, i.e., point-wise model predictions $f_\theta(\dots) \eqqcolon f_\theta^p$ and ground truth $y_\xi(\dots) \eqqcolon y_\xi^p$. We do so via the following metrics.
We define approximation disparity
\[
\text{Approx. disp.}\colon \sqrt{ \sum_{p} \norm{y_\xi^p - f_\theta^p}_2^2 / \sum_{p} \norm{y_\xi^p}_2^2 }
\]
which measures the similarity between the vector fields. Furthermore, we use mean cosine similarity between vectors
\[
\text{Cos. similarity}\colon \mean_{p} \cos \angle (y_\xi^p, f_\theta^p)
\]
which ranges between -1 (opposite)
and 1 (proportional) and measures directional agreement independent of magnitude.

\begin{figure}
	\centering
	\includegraphics{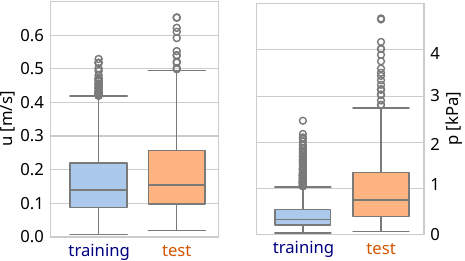}
	\caption{\revision{}{\textbf{Distribution of MAE} $\downarrow$ (over time and subjects) between the machine learning model compared to CFD as boxplots for both training and test split for estimation of velocity $u$ and pressure $p$, respectively. The backend model was LaB-GATr.}}\label{fig:distribution}
\end{figure}

\begin{figure*}
	\centering
	\includegraphics{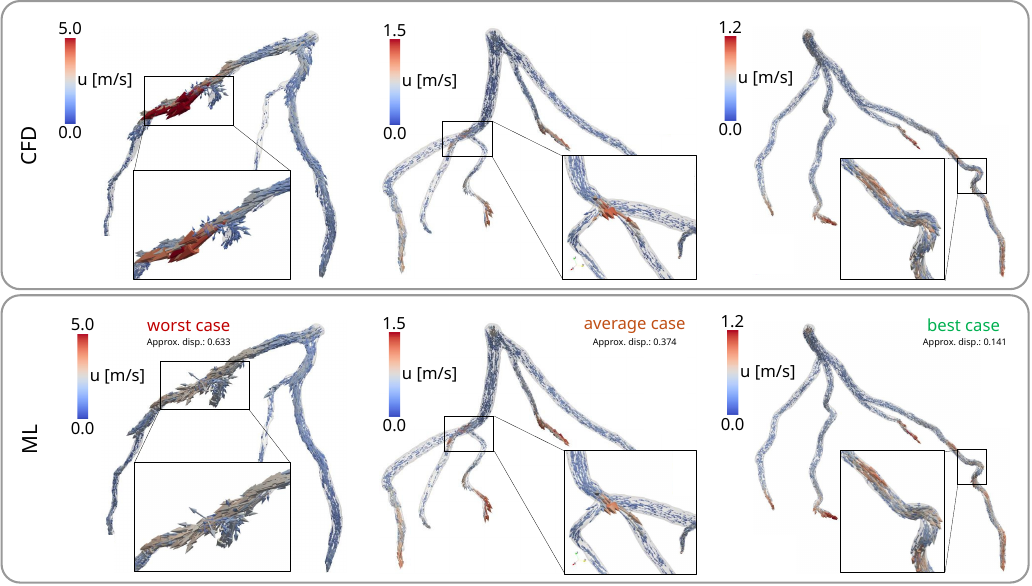}
	\caption{\revision{}{\textbf{Comparative study} of different levels of accuracy (w.r.t. Approx. disp. associated with $u$) of the machine learning (ML) model in comparison to CFD in different test subjects (generalisation). We visualise velocity field vectors corresponding to isolated points in time. The backend model was LaB-GATr.}}\label{fig:podium}
\end{figure*}

\section{Experiments and results}
\subsection{Implementation}
We implemented all neural networks in Python using PyTorch~\citep{PaszkeGross2019} and PyTorch Geometric~\citep{FeyLenssen2019}. For PointNet++, we opted for an aggressive sampling strategy with cumulative factors of 0.0033, 0.3344, 0.6656 and 0.9967 to account for the large-scale data. Likewise, we chose a compression ratio of 0.0033 for LaB-GATr and LaB-VaTr. We let LaB-GATr use cross-attention for the compression module and LaB-VaTr message passing. We chose these settings since they resulted in favourable runtime and accuracy. We chose hidden channels size so all networks had approximately 0.65 million trainable parameters.
We used PyTorch just-in-time compilation to speed up training for MLP while the other models did not benefit from it due to unsupported software.

\subsection{Training details}
We split the 74 patient-specific CFD simulations into 55 training, 4 validation and 15 test cases. We ensured that training and test split had the same ratio between left and right coronary arteries. Left and right coronary artery of the same subject were considered separate cases.
Where the dataset contained geometries of the left as well as the right coronary artery of the same subject, both where included in the same dataset split. All models were trained with batch size 12 (enabled by gradient accumulation) on NVIDIA L40 (48 GB) GPUs using Adam optimiser (learning rate $3 \cdot 10^{-4}$). We trained MLP, PointNet++ and Lab-VaTr each for 1000 epochs on a single GPU which took 56:28 h on average.
We trained LaB-GATr for 200 epochs (47:00 h) on four parallel GPUs to ensure fair comparison. For each neural network, we performed two training runs, with and without (both) exponential learning rate decay
and gradient clipping. LAB-GATr showed increased accuracy while the other models did not benefit from it. Our implementation is publicly available.\footnote{\href{https://github.com/sukjulian/deep-vectorised-operators}{github.com/sukjulian/deep-vectorised-operators}}

\begin{table}
	\caption{\revision{}{\textbf{Mean absolute error (MAE)} $\downarrow$ of the machine learning model compared to CFD. We associate each time step in each test subject with a scalar MAE value and report the average and maximum among all of these. The backend model was LaB-GATr.}}\label{tab:mae}
	\centering
	\renewcommand{\arraystretch}{1.3}
	\begin{tabular}{@{}cccccc@{}}
		\toprule
		\multicolumn{2}{c}{$u$ [m/s]} & \multicolumn{2}{c}{$p$ [kPa]} & \multicolumn{2}{c}{vFFR}\\
		\cmidrule(r){1-2} \cmidrule(lr){3-4} \cmidrule(l){5-6}
		Average & Max. & Average & Max. & Average & Max.\\
		\midrule
		0.189 & 0.653 & 1.0 & 4.7 & 0.07 & 0.25 \\
		\bottomrule
	\end{tabular}
\end{table}

\begin{figure*}
	\centering
	\includegraphics{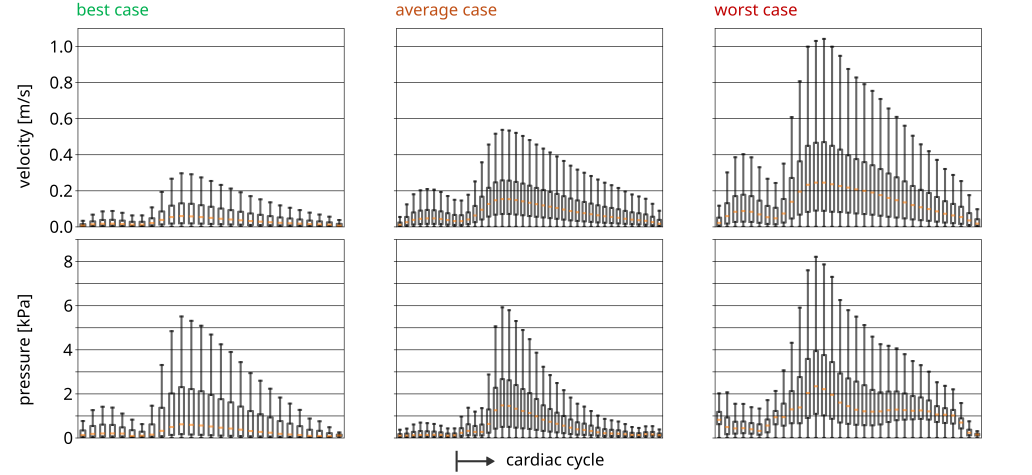}
	\caption{\revision{}{\textbf{MAE $\downarrow$ over cardiac cycle} for test subjects (generalisation) corresponding to different levels of accuracy (w.r.t time-averaged MAE associated with $u$). Distribution of MAE over spatial points in the artery is shown as boxplot for each temporal point in the cardiac cycle. We exclude outliers for visual clarity (due to the very large number of spatial points). The backend model was LaB-GATr.}}\label{fig:spacetime}
\end{figure*}

\subsection{Hemodynamics estimation}
Figure~\ref{fig:qualitative} compares the estimated pulsatile hemodynamics by our deep vectorised operator $G_{\gamma, \xi}$  in a \revision{patient}{subject} from the test split to CFD. Here, $G_{\gamma, \xi}$ uses LaB-GATr as backend. From the streamlines, we see good overall agreement between ground truth and prediction. We observe the highest disparity in regions with high-frequency flow components\footnote{\revision{}{By high-frequency, we mean high variation in direction and magnitude on a relatively small spatial and temporal scale.}}, e.g., around the bifurcation and along the tortuous downstream vasculature during peak systole (time steps four and five). Looking at the pressure maps, we observe excellent agreement between $G_{\gamma, \xi}$ and CFD in upstream vasculature which slightly and gradually decreases moving downstream (compare again peak systole). This might be because we prescribe boundary conditions $\xi$ exclusively at the artery inlet in $G_{\gamma, \xi}$, since it is easier to measure in a clinical scenario. \revision{}{We put the qualitative evaluation into perspective by reporting mean absolute error (MAE) between estimation and ground truth across the test split in Table~\ref{tab:mae}. We find that MAE is an order of magnitude lower than the values observed in  Figure~\ref{fig:qualitative}. Furthermore, in Figure~\ref{fig:distribution} we provide comparison of the distribution of MAE across both training and test split. Figure~\ref{fig:podium} shows three levels of accuracy -- worst, average and best -- among time steps in all subjects. We chose velocity $u$ for this comparison. In the average case, we find good agreement, even for difficult regions around the bifurcations. In the worst case, which is associated with peak systole, we observe difficulties capturing the high-frequency\minorrevision{, turbulent}{} flow components which supports the observations above. In Figure~\ref{fig:spacetime} we present the spatial distribution of MAE at each point in the cardiac cycle for three subjects in the test set, corresponding to three levels of accuracy w.r.t. estimation of velocity $u$. We observe that the highest errors, both for velocity and pressure estimation, occur during peak systole in all three cases. The amplitude of the boxplots can be interpreted as spatial deviation of MAE. For example, the error is higher in arterial bifurcations (compare Figure~\ref{fig:podium}). We observe that during peak systole, spatial deviation of MAE increases, which can be explained by formation of high-frequency flow components.}

\revision{}{\subsection{Estimated fractional flow reserve}
We evaluate the capability of our machine learning model to estimate ``virtual" fractional flow reserve (vFFR), which we define as pressure field $p$, divided point-wise by the average pressure over the artery inlet. In Table~\ref{tab:mae} we compare estimated vFFR between our deep vectorised operator and CFD on the basis of MAE across the test split. We find that average MAE is as low as 0.07 while maximum MAE among all time steps in all subjects breaches the clinical decision threshold (0.75 - 0.8) at 0.25. Note that this evaluation is based on the volumetric pressure field rather than a surface map (the latter of which is sufficient for vFFR) and no task-specific fine-tuning was performed.}

\subsection{Backend study}
In Table~\ref{tab:backend} we compare model predictions on the held-out test split to the corresponding ground truth values via Approx. disp. and Cos. similarity. LaB-GATr achieves the highest accuracy on the estimation of $u \concat p$, followed by PointNet++, LaB-VaTr and MLP with statistically significant ($p < 0.05$ in a one-way ANOVA test) difference. When isolating the target fields, PointNet++ and LaB-GATr are tied for most accurate (directional) velocity prediction in terms of Cos. similarity while LaB-GATr achieves lower Approx. disp.. LaB-VaTr achieves the lowest mean Approx. disp. in pressure field estimation, however, there is no statistically significant difference to LaB-GATr. In terms of training complexity, PointNet++ is the fastest at 164 seconds per epoch, outpacing MLP despite just-in-time compilation. LaB-GATr is computationally heavy to train, requiring 857 seconds per epoch even when parallelised across four GPUs.

\begin{table}
	\caption{\textbf{Multitask versus isolated training target.} We re-trained PointNet++ three times to predict only velocity $u$, only pressure $p$ and both $u \concat p$ jointly and report the resulting metrics.}\label{tab:joint}
	\centering
	\renewcommand{\arraystretch}{1.3}
	\resizebox{\columnwidth}{!}{
		\begin{tabular}{@{}cccc@{}}
			\toprule
			& \multicolumn{2}{c}{$u$} & $p$\\
			\cmidrule(lr){2-3} \cmidrule(lr){4-4}
			Training target & Approx. disp. $\downarrow$ & Cos. similarity $\uparrow$ & Approx. disp. $\downarrow$\\
			\midrule
			$u \concat p$ & $\mathbf{0.386} \pm 0.072$ & $\mathbf{\hphantom{-}0.76} \pm 0.04$ & $\mathbf{0.114} \pm 0.073$\\
			$u$ & $\mathbf{0.388} \pm 0.074$ & $\mathbf{\hphantom{-}0.76} \pm 0.04$ & $0.345 \pm 0.148$\\
			$p$ & $1.006 \pm 0.018$ & $-0.16 \pm 0.12$ & $\mathbf{0.113} \pm 0.068$\\
			\bottomrule
		\end{tabular}
	}
\end{table}

\subsection{Multitask versus isolated target learning}
We investigate the possible advantage of isolating the estimation of velocity and pressure into separately trained neural networks. To this end, we re-trained a model with the PointNet++ backend for 400 epochs three times, with prediction target $u \concat p$ (jointly), as well as $u$ and $p$ (isolated). In Table~\ref{tab:joint} we present the results. We did not find statistically significant ($p < 0.05$ in a one-way ANOVA test) difference between multitask and isolated target learning, indicating that multitask learning neither benefits nor harms prediction of velocity or pressure fields.

\begin{figure*}
	\centering
	\includegraphics{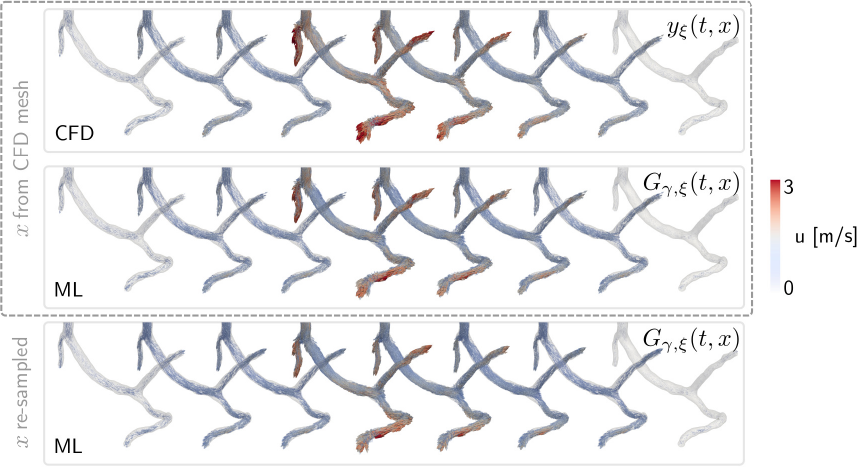}
	\caption{\textbf{Discretisation sensitivity} analysis of machine learning (ML) model in a test subject (generalisation). We evaluate $G_{\gamma, \xi}(t, x)$ on positions $x$ induced by the CFD mesh as well randomly sampled positions from the domain and visualise the resulting velocity fields. The backend model was LaB-GATr.}\label{fig:discretisation}
\end{figure*}

\subsection{Sensitivity to discretisation}
We investigate our models' capability for zero-shot generalisation to re-discretisation of the domain. To this end, we discard the mesh vertices and sample new positions (freely in space), approximately evenly from the surface of the artery and from within the artery lumen. We make sure the number of sampled surface and volume points are approximately equal to the CFD meshes used during training. We then evaluate the pre-trained  (vectorised) conditional neural field $f_\theta$ with MLP, PointNet++, LaB-GATr and LaB-VaTr as backends on these newly sampled points. To enable comparison to the ground truth, we perform proportional interpolation of steady state and pulsatile hemodynamic fields $\gamma, y_\xi$ based on distance to four closest points in the source mesh.
In Table~\ref{tab:discretisation} we report $\Delta$ Approx. disp. (the difference in Approx. disp.) between training resolution and randomly sampled points in subjects from the test split. Positive values mean that Approx. disp. increased, i.e., accuracy decreased. We find that for all four backends $\Delta$ Approx. disp. is marginal. Furthermore, we find that for MLP, LaB-GATr and LaB-VaTr, the difference is not statistically significant ($p > 0.05$ in a one-way ANOVA test), i.e., re-discretisation has not changed their accuracy. The PointNet++ backend is the only one resulting in statistically significant $\Delta$ Approx. disp..
Figure.~\ref{fig:discretisation} showcases the discretisation-robustness of (vectorised) conditional neural field $f_\theta = G_{\gamma, \xi}$ for the example of velocity field estimation. Here, $f_\theta$ uses LaB-GATr as backend. We observe that the vector fields resulting from evaluation of $G_{\gamma, \xi}(t, x)$ on positions $x$ from the CFD mesh (training discretisation) and randomly sampled positions visually coincide. The amount of outliers is negligible.

\begin{table}
	\caption{\textbf{Discretisation sensitivity.}
		We evaluate the pre-trained models on new point coordinates that are randomly sampled from the domain. We report the difference $\Delta$ in Approx. disp. for joint prediction of $u \concat p$ and check whether the difference is statistically significant (one-way ANOVA test).}\label{tab:discretisation}
	\centering
	\renewcommand{\arraystretch}{1.3}
	\begin{tabular}{@{}lc@{}}
		\toprule
		& $u \concat p$\\
		\cmidrule(lr){2-2}
		Backend & $\Delta$ Approx. disp. $\downarrow$\\
		\midrule
		MLP & $-0.004\hphantom{^*}$\\
		PointNet++ & $+0.021^*$\\
		LaB-GATr & $-0.000\hphantom{^*}$\\  
		LaB-VaTr & $-0.003\hphantom{^*}$\\
		\bottomrule
		\multicolumn{2}{l}{$^*$significant ($p < 0.05$)}
	\end{tabular}
\end{table}

\section{Discussion and conclusion}
In this work, we presented a novel deep learning approach that takes the end-diastolic steady-state hemodynamics in the coronary arteries and estimates the corresponding pulsatile hemodynamics, over the full cardiac cycle. This idea is based on the observation that steady-state simulations are orders of magnitude faster than pulsatile simulations while \revision{many}{} medically relevant biomarkers \revision{are}{may be} related to pulsatile flow, e.g., multi-directional wall shear stress~\revision{}{\citep{HoogendoornKok2019}}. Our model is conditioned on the patient-specific pulsatile waveform, as well as the flow and pressure over the artery inlet. All of these quantities are easily accessible from direct or indirect clinical measurements.  
We evaluated our method on a dataset of CFD-based hemodynamic simulations in stenotic left and right coronary arteries encompassing 74 cases across a cohort of real-life patients. By casting the learning objective as operator learning from steady-state to pulsatile flow, our method is able to accurately capture the 4D nature of blood flow based on limited amounts of (fairly diverse) data. Steady-state simulations were 24 to 288 times faster than pulsatile simulations in these subjects. Consequently, our approach could potentially be useful in the clinic to make pulsatile hemodynamic fields available in settings where time is critical or computational resources are limited. Medical decision markers like wall shear stress, oscillatory shear index and fractional flow reserve can then be computed from pressure and velocity fields. Importantly, our method is robust w.r.t. discretisation of the flow domain which could be beneficial for inter-operator variability.

We described this approach in a unified framework which we called deep vectorised operator, since our model maps between infinite-dimensional function spaces.
Deep vectorised operators are parametrised by (vectorised) conditional neural fields~\citep{HagnbergerKalimuthu2024} that can be embodied by several backends, as long as they are point-wise permutation-equivariant like message passing or self-attention models. Importantly, \textit{vectorised} neural fields differ from neural fields in the sense that all query points are fed to the model at once, in a vectorised way. Our empirical evidence justifies the nomenclature ``neural field'' by suggesting that deep vectorised operators are discretisation-independent, i.e., enable zero-shot generalisation to re-sampling of the domain. This is remarkable, because the neural architectures we tested, in particular PointNet++~\citep{QiYi2017}, are based on cross-conditioning between the query points. Our evidence suggests that they nevertheless function as a mapping between domain and co-domain, just like traditional neural fields~\citep{XieTakikawa2022} parametrised by point-wise neural architecures like MLP. As expected, the deep vectorised operator with MLP as backend was agnostic to re-sampling of the domain while PointNet++ exhibited small (but statistically significant) sensitivity, hinting at pronounced cross-conditioning.

\revision{}{As data-driven approach, the accuracy of the presented method is proportional to the amount of available data. In our evaluations, we observed loss of accuracy for high-frequency flow components. This indicates that the model is not able to reliably overcome spectral bias~\citep{RahamanBaratin2019} based on the available amount of data. However, we expect improvements when more training data is available. While the presented method maps between function spaces and is thus agnostic to the topology (e.g. number of bifurcations) and discretisation (e.g. number of mesh vertices) of the domain, its estimations will only be accurate insofar the training data sufficiently resembles the data seen at inference time.\footnote{\revision{}{In particular, this means that for clinical adoption, each hospital would have to train its own model on proprietary data or a large representative dataset would have to be collected to train a universal foundation model.}} This argument extends to the encoding of Dirichlet boundary conditions and waveform. The operator learning setup affords representational freedom, e.g., discretisation of any (including real-world) waveform at a fixed number of points in time. Since our training data was based on a scaled template waveform from the literature, we opted for simple statistical measures, though. Ultimately, more sophisticated waveforms would require a larger amount of representative training data, of leveraging which -- we argue -- our model is a-priori capable. Investigating applicability and adaptability to realistic, real-world waveforms is an important avenue for future work.}

Our approach is similar in scope to DeepONet and MIONet as we are learning a deep operator between function spaces. In comparison to MIONet, we sacrificed the entry-wise Hadamard-product structure for scalability and ease of implementation. This affects the learned operator and in turn the available mathematical approximation theory, investigating which is an interesting avenue for future work. In comparison to \cite{MaulZinn2023}, our work explores the incorporation of steady-state hemodynamics as powerful prior for pulsatile hemodynamics, our model returns velocity \textit{and} pressure and our dataset is composed of real-life patients. For future work, we plan to incorporate a balance term for inlet and outlet flux in the training loss \revision{to improve estimations at the artery outlets}{}, as well as continuity and momentum terms\revision{}{, to improve consistency with the flow physics and pressure estimation at the artery outlet}. \revision{Furthermore, we will investigate the quality of extracted medical decision markers like wall shear stress, oscillatory shear index and fractional flow reserve computed from deep vectorised operators' estimated hemodynamic fields.}{A key limitation in the present study is that, while medical decision markers like wall shear stress, oscillatory shear index and (virtual) fractional flow reserve can indeed be computed from deep vectorised operators' estimated velocity and pressure fields, we only showcase this for vFFR. Since wall shear stress (and consequently oscillatory shear index) is a first-order differential quantity which depends on the Jacobian of the velocity field, there is no naive way of computing it and we require interpolation, finite differences or finite elements. In order for deep vectorised operators to output sufficiently smooth velocity fields, we conjecture that at least first-order regularisation is required during training. In detail, we aim to include a Neumann boundary condition relating the near-wall velocity field with the (finite-elements-based) wall shear stress from CFD. As this approach requires differentiation, it blends in with the aforementioned continuity and momentum regularisation. We plan on investigating these jointly.}

In conclusion, deep vectorised operators can be used to estimate pulsatile from steady-state hemodynamics in coronary arteries while being independent of the discretisation of the domain. Furthermore, deep vectorised operators are a useful modelling tool with many expected applications in cardiovascular medicine and biomedical engineering.

\appendix

\section*{Declaration of interests}
The authors declare that they have no known competing financial interests or personal relationships that could have appeared to influence the work reported in this paper.

\section*{Acknowledgements}
This work is funded in part by the 4TU Precision Medicine programme supported by High Tech for a Sustainable Future, a framework commissioned by the four Universities of Technology of the Netherlands. Jelmer M. Wolterink was supported by the NWO domain Applied and Engineering Sciences VENI grant (18192).


\bibliographystyle{cas-model2-names}

\bibliography{cas-refs}  

\begin{thebibliography}{50}
\expandafter\ifx\csname natexlab\endcsname\relax\def\natexlab#1{#1}\fi
\providecommand{\url}[1]{\texttt{#1}}
\providecommand{\href}[2]{#2}
\providecommand{\path}[1]{#1}
\providecommand{\DOIprefix}{doi:}
\providecommand{\ArXivprefix}{arXiv:}
\providecommand{\URLprefix}{URL: }
\providecommand{\Pubmedprefix}{pmid:}
\providecommand{\doi}[1]{\href{http://dx.doi.org/#1}{\path{#1}}}
\providecommand{\Pubmed}[1]{\href{pmid:#1}{\path{#1}}}
\providecommand{\bibinfo}[2]{#2}
\ifx\xfnm\relax \def\xfnm[#1]{\unskip,\space#1}\fi
\bibitem[{Alkin et~al.(2024)Alkin, F{\"{u}}rst, Schmid, Gruber, Holzleitner and
  Brandstetter}]{AlkinFürst2024}
\bibinfo{author}{Alkin, B.}, \bibinfo{author}{F{\"{u}}rst, A.},
  \bibinfo{author}{Schmid, S.}, \bibinfo{author}{Gruber, L.},
  \bibinfo{author}{Holzleitner, M.}, \bibinfo{author}{Brandstetter, J.},
  \bibinfo{year}{2024}.
\newblock \bibinfo{title}{Universal physics transformers: {A} framework for
  efficiently scaling neural operators}, in: \bibinfo{editor}{Globersons, A.},
  \bibinfo{editor}{Mackey, L.}, \bibinfo{editor}{Belgrave, D.},
  \bibinfo{editor}{Fan, A.}, \bibinfo{editor}{Paquet, U.},
  \bibinfo{editor}{Tomczak, J.M.}, \bibinfo{editor}{Zhang, C.} (Eds.),
  \bibinfo{booktitle}{Advances in Neural Information Processing Systems 38:
  Annual Conference on Neural Information Processing Systems 2024, NeurIPS
  2024, Vancouver, BC, Canada, December 10 - 15, 2024}, pp.
  \bibinfo{pages}{0--16}.
\bibitem[{Candreva et~al.(2022)Candreva, Pagnoni, Rizzini, Mizukami, Gallinoro,
  Mazzi, Gallo, Meier, Shinke, Aben, Nagumo, Sonck, Munhoz, Fournier, Barbato,
  Heggermont, Cook, Chiastra, Morbiducci, {De Bruyne}, Muller and
  Collet}]{CandrevaPagnoni2022}
\bibinfo{author}{Candreva, A.}, \bibinfo{author}{Pagnoni, M.},
  \bibinfo{author}{Rizzini, M.L.}, \bibinfo{author}{Mizukami, T.},
  \bibinfo{author}{Gallinoro, E.}, \bibinfo{author}{Mazzi, V.},
  \bibinfo{author}{Gallo, D.}, \bibinfo{author}{Meier, D.},
  \bibinfo{author}{Shinke, T.}, \bibinfo{author}{Aben, J.P.},
  \bibinfo{author}{Nagumo, S.}, \bibinfo{author}{Sonck, J.},
  \bibinfo{author}{Munhoz, D.}, \bibinfo{author}{Fournier, S.},
  \bibinfo{author}{Barbato, E.}, \bibinfo{author}{Heggermont, W.},
  \bibinfo{author}{Cook, S.}, \bibinfo{author}{Chiastra, C.},
  \bibinfo{author}{Morbiducci, U.}, \bibinfo{author}{{De Bruyne}, B.},
  \bibinfo{author}{Muller, O.}, \bibinfo{author}{Collet, C.},
  \bibinfo{year}{2022}.
\newblock \bibinfo{title}{Risk of myocardial infarction based on endothelial
  shear stress analysis using coronary angiography}.
\newblock \bibinfo{journal}{Atherosclerosis} \bibinfo{volume}{342},
  \bibinfo{pages}{28--35}.
\bibitem[{Chung and Cebral(2015)}]{ChungCebral2015}
\bibinfo{author}{Chung, B.}, \bibinfo{author}{Cebral, J.R.},
  \bibinfo{year}{2015}.
\newblock \bibinfo{title}{{CFD} for evaluation and treatment planning of
  aneurysms: review of proposed clinical uses and their challenges}.
\newblock \bibinfo{journal}{Annals of Biomedical Engineering}
  \bibinfo{volume}{43}, \bibinfo{pages}{122--138}.
\bibitem[{Crane et~al.(2017)Crane, Weischedel and
  Wardetzky}]{CraneWeischedel2017}
\bibinfo{author}{Crane, K.}, \bibinfo{author}{Weischedel, C.},
  \bibinfo{author}{Wardetzky, M.}, \bibinfo{year}{2017}.
\newblock \bibinfo{title}{The heat method for distance computation}.
\newblock \bibinfo{journal}{Commun. ACM} \bibinfo{volume}{60},
  \bibinfo{pages}{90--99}.
\bibitem[{Driessen et~al.(2019)Driessen, Danad, Stuijfzand, Raijmakers,
  Schumacher, {van Diemen}, Leipsic, Knuuti, Underwood, {van de Ven}, {van
  Rossum}, Taylor and Knaapen}]{DriessenDanad2019}
\bibinfo{author}{Driessen, R.S.}, \bibinfo{author}{Danad, I.},
  \bibinfo{author}{Stuijfzand, W.J.}, \bibinfo{author}{Raijmakers, P.G.},
  \bibinfo{author}{Schumacher, S.P.}, \bibinfo{author}{{van Diemen}, P.A.},
  \bibinfo{author}{Leipsic, J.A.}, \bibinfo{author}{Knuuti, J.},
  \bibinfo{author}{Underwood, S.R.}, \bibinfo{author}{{van de Ven}, P.M.},
  \bibinfo{author}{{van Rossum}, A.C.}, \bibinfo{author}{Taylor, C.A.},
  \bibinfo{author}{Knaapen, P.}, \bibinfo{year}{2019}.
\newblock \bibinfo{title}{Comparison of coronary computed tomography
  angiography, fractional flow reserve, and perfusion imaging for ischemia
  diagnosis}.
\newblock \bibinfo{journal}{Journal of the American College of Cardiology}
  \bibinfo{volume}{73}, \bibinfo{pages}{161--173}.
\bibitem[{Fathi et~al.(2020)Fathi, Perez-Raya, Baghaie, Berg, Janiga, Arzani
  and D’Souza}]{FathiPerezRaya2020}
\bibinfo{author}{Fathi, M.F.}, \bibinfo{author}{Perez-Raya, I.},
  \bibinfo{author}{Baghaie, A.}, \bibinfo{author}{Berg, P.},
  \bibinfo{author}{Janiga, G.}, \bibinfo{author}{Arzani, A.},
  \bibinfo{author}{D’Souza, R.M.}, \bibinfo{year}{2020}.
\newblock \bibinfo{title}{Super-resolution and denoising of {4D}-flow mri using
  physics-informed deep neural nets}.
\newblock \bibinfo{journal}{Computer Methods and Programs in Biomedicine}
  \bibinfo{volume}{197}, \bibinfo{pages}{105729}.
\bibitem[{Fedorov et~al.(2012)Fedorov, Beichel, Kalpathy-Cramer, Finet,
  Fillion-Robin, Pujol, Bauer, Jennings, Fennessy, Sonka, Buatti, Aylward,
  Miller, Pieper and Kikinis}]{Fedorov2012}
\bibinfo{author}{Fedorov, A.}, \bibinfo{author}{Beichel, R.},
  \bibinfo{author}{Kalpathy-Cramer, J.}, \bibinfo{author}{Finet, J.},
  \bibinfo{author}{Fillion-Robin, J.C.}, \bibinfo{author}{Pujol, S.},
  \bibinfo{author}{Bauer, C.}, \bibinfo{author}{Jennings, D.},
  \bibinfo{author}{Fennessy, F.}, \bibinfo{author}{Sonka, M.},
  \bibinfo{author}{Buatti, J.}, \bibinfo{author}{Aylward, S.},
  \bibinfo{author}{Miller, J.V.}, \bibinfo{author}{Pieper, S.},
  \bibinfo{author}{Kikinis, R.}, \bibinfo{year}{2012}.
\newblock \bibinfo{title}{{3D Slicer as an image computing platform for the
  Quantitative Imaging Network}}.
\newblock \bibinfo{journal}{Magnetic resonance imaging} \bibinfo{volume}{30},
  \bibinfo{pages}{1323–1341}.
\bibitem[{Fey and Lenssen(2019)}]{FeyLenssen2019}
\bibinfo{author}{Fey, M.}, \bibinfo{author}{Lenssen, J.E.},
  \bibinfo{year}{2019}.
\newblock \bibinfo{title}{Fast graph representation learning with {PyTorch
  Geometric}}, in: \bibinfo{booktitle}{ICLR Workshop on Representation Learning
  on Graphs and Manifolds}, p. \bibinfo{pages}{0–9}.
\bibitem[{Flemister et~al.(2020)Flemister, Hatoum, Guhan, Zebhi, Lincoln and
  Crestanello}]{Flemister2020}
\bibinfo{author}{Flemister, D.C.}, \bibinfo{author}{Hatoum, H.},
  \bibinfo{author}{Guhan, V.}, \bibinfo{author}{Zebhi, B.},
  \bibinfo{author}{Lincoln, J.}, \bibinfo{author}{Crestanello, Juan~Dasi,
  L.P.}, \bibinfo{year}{2020}.
\newblock \bibinfo{title}{{Effect of Left and Right Coronary Flow Waveforms on
  Aortic Sinus Hemodynamics and Leaflet Shear Stress: Correlation with
  Calcification Locations}}.
\newblock \bibinfo{journal}{Annals of Biomedical Engineering}
  \bibinfo{volume}{48}, \bibinfo{pages}{2796--2808}.
\bibitem[{Gharleghi et~al.(2022)Gharleghi, Sowmya and
  Beier}]{GharleghiSowmya2022}
\bibinfo{author}{Gharleghi, R.}, \bibinfo{author}{Sowmya, A.},
  \bibinfo{author}{Beier, S.}, \bibinfo{year}{2022}.
\newblock \bibinfo{title}{Transient wall shear stress estimation in coronary
  bifurcations using convolutional neural networks}.
\newblock \bibinfo{journal}{Computer Methods and Programs in Biomedicine}
  \bibinfo{volume}{225}, \bibinfo{pages}{107013}.
\bibitem[{Gropp et~al.(2020)Gropp, Yariv, Haim, Atzmon and
  Lipman}]{GroppYariv2020}
\bibinfo{author}{Gropp, A.}, \bibinfo{author}{Yariv, L.},
  \bibinfo{author}{Haim, N.}, \bibinfo{author}{Atzmon, M.},
  \bibinfo{author}{Lipman, Y.}, \bibinfo{year}{2020}.
\newblock \bibinfo{title}{Implicit geometric regularization for learning
  shapes}, in: \bibinfo{booktitle}{Proceedings of the 37th International
  Conference on Machine Learning}, \bibinfo{publisher}{JMLR.org}. pp.
  \bibinfo{pages}{0--11}.
\bibitem[{Hagnberger et~al.(2024)Hagnberger, Kalimuthu, Musekamp and
  Niepert}]{HagnbergerKalimuthu2024}
\bibinfo{author}{Hagnberger, J.}, \bibinfo{author}{Kalimuthu, M.},
  \bibinfo{author}{Musekamp, D.}, \bibinfo{author}{Niepert, M.},
  \bibinfo{year}{2024}.
\newblock \bibinfo{title}{Vectorized conditional neural fields: {A} framework
  for solving time-dependent parametric partial differential equations}, in:
  \bibinfo{booktitle}{Forty-first International Conference on Machine Learning,
  {ICML} 2024, Vienna, Austria, July 21-27, 2024},
  \bibinfo{publisher}{OpenReview.net}. pp. \bibinfo{pages}{0--35}.
\bibitem[{Hoogendoorn et~al.(2019)Hoogendoorn, Kok, Hartman, de~Nisco,
  Casadonte, Chiastra, Coenen, Korteland, Van~der Heiden, Gijsen, Duncker,
  van~der Steen and Wentzel}]{HoogendoornKok2019}
\bibinfo{author}{Hoogendoorn, A.}, \bibinfo{author}{Kok, A.M.},
  \bibinfo{author}{Hartman, E.M.J.}, \bibinfo{author}{de~Nisco, G.},
  \bibinfo{author}{Casadonte, L.}, \bibinfo{author}{Chiastra, C.},
  \bibinfo{author}{Coenen, A.}, \bibinfo{author}{Korteland, S.A.},
  \bibinfo{author}{Van~der Heiden, K.}, \bibinfo{author}{Gijsen, F.J.H.},
  \bibinfo{author}{Duncker, D.J.}, \bibinfo{author}{van~der Steen, A.F.W.},
  \bibinfo{author}{Wentzel, J.J.}, \bibinfo{year}{2019}.
\newblock \bibinfo{title}{Multidirectional wall shear stress promotes advanced
  coronary plaque development: comparing five shear stress metrics}.
\newblock \bibinfo{journal}{Cardiovascular Research} \bibinfo{volume}{116},
  \bibinfo{pages}{1136--1146}.
\bibitem[{Jin et~al.(2022)Jin, Meng and Lu}]{JinMeng2022}
\bibinfo{author}{Jin, P.}, \bibinfo{author}{Meng, S.}, \bibinfo{author}{Lu,
  L.}, \bibinfo{year}{2022}.
\newblock \bibinfo{title}{Mionet: Learning multiple-input operators via tensor
  product}.
\newblock \bibinfo{journal}{SIAM Journal on Scientific Computing}
  \bibinfo{volume}{44}, \bibinfo{pages}{A3490--A3514}.
\bibitem[{{Kontogiannis} and {Juniper}(2021)}]{KontogiannisJuniper2021}
\bibinfo{author}{{Kontogiannis}, A.}, \bibinfo{author}{{Juniper}, M.},
  \bibinfo{year}{2021}.
\newblock \bibinfo{title}{{Physics-informed compressed sensing: reconstruction
  of magnetic resonance velocimetry signals as an inverse Navier-Stokes
  problem}}, in: \bibinfo{booktitle}{APS Division of Fluid Dynamics Meeting
  Abstracts}, p. \bibinfo{pages}{H20.001}.
\bibitem[{Kovachki et~al.(2024)Kovachki, Li, Liu, Azizzadenesheli,
  Bhattacharya, Stuart and Anandkumar}]{KovachkiLi2024}
\bibinfo{author}{Kovachki, N.}, \bibinfo{author}{Li, Z.}, \bibinfo{author}{Liu,
  B.}, \bibinfo{author}{Azizzadenesheli, K.}, \bibinfo{author}{Bhattacharya,
  K.}, \bibinfo{author}{Stuart, A.}, \bibinfo{author}{Anandkumar, A.},
  \bibinfo{year}{2024}.
\newblock \bibinfo{title}{Neural operator: learning maps between function
  spaces with applications to pdes}.
\newblock \bibinfo{journal}{J. Mach. Learn. Res.} \bibinfo{volume}{24}.
\bibitem[{Lee et~al.(2019)Lee, Choi, Koo, Hwang, Park, Zhang, Kim, Tong, Kim,
  Grady, Doh, Nam, Shin, Cho, Choi, Chun, Choi, Nørgaard, Christiansen,
  Niemen, Otake, Penicka, {de Bruyne}, Kubo, Akasaka, Narula, Douglas, Taylor
  and Kim}]{LeeChoi2019}
\bibinfo{author}{Lee, J.M.}, \bibinfo{author}{Choi, G.}, \bibinfo{author}{Koo,
  B.K.}, \bibinfo{author}{Hwang, D.}, \bibinfo{author}{Park, J.},
  \bibinfo{author}{Zhang, J.}, \bibinfo{author}{Kim, K.J.},
  \bibinfo{author}{Tong, Y.}, \bibinfo{author}{Kim, H.J.},
  \bibinfo{author}{Grady, L.}, \bibinfo{author}{Doh, J.H.},
  \bibinfo{author}{Nam, C.W.}, \bibinfo{author}{Shin, E.S.},
  \bibinfo{author}{Cho, Y.S.}, \bibinfo{author}{Choi, S.Y.},
  \bibinfo{author}{Chun, E.J.}, \bibinfo{author}{Choi, J.H.},
  \bibinfo{author}{Nørgaard, B.L.}, \bibinfo{author}{Christiansen, E.H.},
  \bibinfo{author}{Niemen, K.}, \bibinfo{author}{Otake, H.},
  \bibinfo{author}{Penicka, M.}, \bibinfo{author}{{de Bruyne}, B.},
  \bibinfo{author}{Kubo, T.}, \bibinfo{author}{Akasaka, T.},
  \bibinfo{author}{Narula, J.}, \bibinfo{author}{Douglas, P.S.},
  \bibinfo{author}{Taylor, C.A.}, \bibinfo{author}{Kim, H.S.},
  \bibinfo{year}{2019}.
\newblock \bibinfo{title}{Identification of high-risk plaques destined to cause
  acute coronary syndrome using coronary computed tomographic angiography and
  computational fluid dynamics}.
\newblock \bibinfo{journal}{JACC: Cardiovascular Imaging} \bibinfo{volume}{12},
  \bibinfo{pages}{1032--1043}.
\bibitem[{Li et~al.(2021a)Li, Wang, Zhang, Tupin, Qiao, Liu, Ohta and
  Anzai}]{LiWang2021}
\bibinfo{author}{Li, G.}, \bibinfo{author}{Wang, H.}, \bibinfo{author}{Zhang,
  M.}, \bibinfo{author}{Tupin, S.}, \bibinfo{author}{Qiao, A.},
  \bibinfo{author}{Liu, Y.}, \bibinfo{author}{Ohta, M.},
  \bibinfo{author}{Anzai, H.}, \bibinfo{year}{2021}a.
\newblock \bibinfo{title}{Prediction of {3D} cardiovascular hemodynamics before
  and after coronary artery bypass surgery via deep learning}.
\newblock \bibinfo{journal}{Communications Biology} \bibinfo{volume}{4}.
\bibitem[{Li et~al.(2021b)Li, Kovachki, Azizzadenesheli, Liu, Bhattacharya,
  Stuart and Anandkumar}]{LiKovachki2021}
\bibinfo{author}{Li, Z.}, \bibinfo{author}{Kovachki, N.B.},
  \bibinfo{author}{Azizzadenesheli, K.}, \bibinfo{author}{Liu, B.},
  \bibinfo{author}{Bhattacharya, K.}, \bibinfo{author}{Stuart, A.M.},
  \bibinfo{author}{Anandkumar, A.}, \bibinfo{year}{2021}b.
\newblock \bibinfo{title}{Fourier neural operator for parametric partial
  differential equations}, in: \bibinfo{booktitle}{9th International Conference
  on Learning Representations, {ICLR} 2021, Virtual Event, Austria, May 3-7,
  2021}, \bibinfo{publisher}{OpenReview.net}. pp. \bibinfo{pages}{0--16}.
\bibitem[{Liang et~al.(2020)Liang, Mao and Sun}]{LiangMao2020}
\bibinfo{author}{Liang, L.}, \bibinfo{author}{Mao, W.}, \bibinfo{author}{Sun,
  W.}, \bibinfo{year}{2020}.
\newblock \bibinfo{title}{A feasibility study of deep learning for predicting
  hemodynamics of human thoracic aorta}.
\newblock \bibinfo{journal}{Journal of Biomechanics} \bibinfo{volume}{99},
  \bibinfo{pages}{109544}.
\bibitem[{Lopes et~al.(2020)Lopes, Puga, Teixeira and Lima}]{LopesPuga2020}
\bibinfo{author}{Lopes, D.}, \bibinfo{author}{Puga, H.},
  \bibinfo{author}{Teixeira, J.}, \bibinfo{author}{Lima, R.},
  \bibinfo{year}{2020}.
\newblock \bibinfo{title}{Blood flow simulations in patient-specific geometries
  of the carotid artery: A systematic review}.
\newblock \bibinfo{journal}{Journal of Biomechanics} \bibinfo{volume}{111},
  \bibinfo{pages}{110019}.
\bibitem[{Lu et~al.(2021)Lu, Jin, Pang, Zhang and Karniadakis}]{LuJin2021}
\bibinfo{author}{Lu, L.}, \bibinfo{author}{Jin, P.}, \bibinfo{author}{Pang,
  G.}, \bibinfo{author}{Zhang, Z.}, \bibinfo{author}{Karniadakis, G.E.},
  \bibinfo{year}{2021}.
\newblock \bibinfo{title}{Learning nonlinear operators via {DeepONet} based on
  the universal approximation theorem of operators}.
\newblock \bibinfo{journal}{Nat. Mach. Intell.} \bibinfo{volume}{3},
  \bibinfo{pages}{218--229}.
\bibitem[{Maul et~al.(2023)Maul, Zinn, Wagner, Thies, Rohleder, Pfaff,
  Kowarschik, Birkhold and Maier}]{MaulZinn2023}
\bibinfo{author}{Maul, N.}, \bibinfo{author}{Zinn, K.},
  \bibinfo{author}{Wagner, F.}, \bibinfo{author}{Thies, M.},
  \bibinfo{author}{Rohleder, M.}, \bibinfo{author}{Pfaff, L.},
  \bibinfo{author}{Kowarschik, M.}, \bibinfo{author}{Birkhold, A.},
  \bibinfo{author}{Maier, A.}, \bibinfo{year}{2023}.
\newblock \bibinfo{title}{Transient hemodynamics prediction using an efficient
  octree-based deep learning model}, in: \bibinfo{editor}{Frangi, A.},
  \bibinfo{editor}{de~Bruijne, M.}, \bibinfo{editor}{Wassermann, D.},
  \bibinfo{editor}{Navab, N.} (Eds.), \bibinfo{booktitle}{Information
  Processing in Medical Imaging}, \bibinfo{publisher}{Springer Nature
  Switzerland}, \bibinfo{address}{Cham}. pp. \bibinfo{pages}{183--194}.
\bibitem[{Morales~Ferez et~al.(2021)Morales~Ferez, Mill, Juhl, Acebes, Iriart,
  Legghe, Cochet, De~Backer, Paulsen and Camara}]{MoralesFerezMill2021}
\bibinfo{author}{Morales~Ferez, X.}, \bibinfo{author}{Mill, J.},
  \bibinfo{author}{Juhl, K.A.}, \bibinfo{author}{Acebes, C.},
  \bibinfo{author}{Iriart, X.}, \bibinfo{author}{Legghe, B.},
  \bibinfo{author}{Cochet, H.}, \bibinfo{author}{De~Backer, O.},
  \bibinfo{author}{Paulsen, R.R.}, \bibinfo{author}{Camara, O.},
  \bibinfo{year}{2021}.
\newblock \bibinfo{title}{Deep learning framework for real-time estimation of
  in-silico thrombotic risk indices in the left atrial appendage}.
\newblock \bibinfo{journal}{Frontiers in Physiology} \bibinfo{volume}{12}.
\bibitem[{Nannini et~al.(2024)Nannini, Saitta, Mariani, Maragna, Baggiano,
  Mushtaq, Pontone and Redaelli}]{Nannini2024}
\bibinfo{author}{Nannini, G.}, \bibinfo{author}{Saitta, S.},
  \bibinfo{author}{Mariani, L.}, \bibinfo{author}{Maragna, R.},
  \bibinfo{author}{Baggiano, A.}, \bibinfo{author}{Mushtaq, S.},
  \bibinfo{author}{Pontone, G.}, \bibinfo{author}{Redaelli, A.},
  \bibinfo{year}{2024}.
\newblock \bibinfo{title}{An automated and time-efficient framework for
  simulation of coronary blood flow under steady and pulsatile conditions}.
\newblock \bibinfo{journal}{Computer Methods and Programs in Biomedicine}
  \bibinfo{volume}{257}, \bibinfo{pages}{108415}.
\bibitem[{Park et~al.(2019)Park, Florence, Straub, Newcombe and
  Lovegrove}]{ParkFlorence2019}
\bibinfo{author}{Park, J.}, \bibinfo{author}{Florence, P.},
  \bibinfo{author}{Straub, J.}, \bibinfo{author}{Newcombe, R.},
  \bibinfo{author}{Lovegrove, S.}, \bibinfo{year}{2019}.
\newblock \bibinfo{title}{Deepsdf: Learning continuous signed distance
  functions for shape representation}, in: \bibinfo{booktitle}{2019 IEEE/CVF
  Conference on Computer Vision and Pattern Recognition (CVPR)},
  \bibinfo{publisher}{IEEE Computer Society}, \bibinfo{address}{Los Alamitos,
  CA, USA}. pp. \bibinfo{pages}{165--174}.
\bibitem[{Paszke et~al.(2019)Paszke, Gross, Massa, Lerer, Bradbury, Chanan,
  Killeen, Lin, Gimelshein, Antiga, Desmaison, K\"{o}pf, Yang, DeVito, Raison,
  Tejani, Chilamkurthy, Steiner, Fang, Bai and Chintala}]{PaszkeGross2019}
\bibinfo{author}{Paszke, A.}, \bibinfo{author}{Gross, S.},
  \bibinfo{author}{Massa, F.}, \bibinfo{author}{Lerer, A.},
  \bibinfo{author}{Bradbury, J.}, \bibinfo{author}{Chanan, G.},
  \bibinfo{author}{Killeen, T.}, \bibinfo{author}{Lin, Z.},
  \bibinfo{author}{Gimelshein, N.}, \bibinfo{author}{Antiga, L.},
  \bibinfo{author}{Desmaison, A.}, \bibinfo{author}{K\"{o}pf, A.},
  \bibinfo{author}{Yang, E.}, \bibinfo{author}{DeVito, Z.},
  \bibinfo{author}{Raison, M.}, \bibinfo{author}{Tejani, A.},
  \bibinfo{author}{Chilamkurthy, S.}, \bibinfo{author}{Steiner, B.},
  \bibinfo{author}{Fang, L.}, \bibinfo{author}{Bai, J.},
  \bibinfo{author}{Chintala, S.}, \bibinfo{year}{2019}.
\newblock \bibinfo{title}{{PyTorch}: an imperative style, high-performance deep
  learning library}, in: \bibinfo{booktitle}{Proceedings of the 33rd
  International Conference on Neural Information Processing Systems},
  \bibinfo{publisher}{Curran Associates Inc.}, \bibinfo{address}{Red Hook, NY,
  USA}. p. \bibinfo{pages}{0–12}.
\bibitem[{Pegolotti et~al.(2024)Pegolotti, Pfaller, Rubio, Ding, {Brugarolas
  Brufau}, Darve and Marsden}]{PegolottiPfaller2024}
\bibinfo{author}{Pegolotti, L.}, \bibinfo{author}{Pfaller, M.R.},
  \bibinfo{author}{Rubio, N.L.}, \bibinfo{author}{Ding, K.},
  \bibinfo{author}{{Brugarolas Brufau}, R.}, \bibinfo{author}{Darve, E.},
  \bibinfo{author}{Marsden, A.L.}, \bibinfo{year}{2024}.
\newblock \bibinfo{title}{Learning reduced-order models for cardiovascular
  simulations with graph neural networks}.
\newblock \bibinfo{journal}{Computers in Biology and Medicine}
  \bibinfo{volume}{168}, \bibinfo{pages}{107676}.
\bibitem[{Pontone et~al.(2017)Pontone, Moharem-Elgamal, Maurovich-Horvat,
  Gaemperli, Pugliese, Westwood, Stefanidis, Fox and Popescu}]{Pontone2017}
\bibinfo{author}{Pontone, G.}, \bibinfo{author}{Moharem-Elgamal, S.},
  \bibinfo{author}{Maurovich-Horvat, P.}, \bibinfo{author}{Gaemperli, O.},
  \bibinfo{author}{Pugliese, F.}, \bibinfo{author}{Westwood, M.},
  \bibinfo{author}{Stefanidis, A.}, \bibinfo{author}{Fox, K.F.},
  \bibinfo{author}{Popescu, B.A.}, \bibinfo{year}{2017}.
\newblock \bibinfo{title}{{Training in cardiac computed tomography: EACVI
  certification process}}.
\newblock \bibinfo{journal}{European Heart Journal - Cardiovascular Imaging}
  \bibinfo{volume}{19}, \bibinfo{pages}{123--126}.
\bibitem[{Qi et~al.(2017)Qi, Yi, Su and Guibas}]{QiYi2017}
\bibinfo{author}{Qi, C.R.}, \bibinfo{author}{Yi, L.}, \bibinfo{author}{Su, H.},
  \bibinfo{author}{Guibas, L.J.}, \bibinfo{year}{2017}.
\newblock \bibinfo{title}{{PointNet++}: Deep hierarchical feature learning on
  point sets in a metric space}, in: \bibinfo{editor}{Guyon, I.},
  \bibinfo{editor}{von Luxburg, U.}, \bibinfo{editor}{Bengio, S.},
  \bibinfo{editor}{Wallach, H.M.}, \bibinfo{editor}{Fergus, R.},
  \bibinfo{editor}{Vishwanathan, S.V.N.}, \bibinfo{editor}{Garnett, R.} (Eds.),
  \bibinfo{booktitle}{Advances in Neural Information Processing Systems 30:
  Annual Conference on Neural Information Processing Systems 2017, December
  4-9, 2017, Long Beach, CA, {USA}}, pp. \bibinfo{pages}{5099--5108}.
\bibitem[{Rabe and Staats(2021)}]{RabeStaats}
\bibinfo{author}{Rabe, M.N.}, \bibinfo{author}{Staats, C.},
  \bibinfo{year}{2021}.
\newblock \bibinfo{title}{Self-attention does not need {$O(n^2)$} memory}, in:
  \bibinfo{booktitle}{n/a}, pp. \bibinfo{pages}{1--8}.
\bibitem[{Rahaman et~al.(2019)Rahaman, Baratin, Arpit, Draxler, Lin, Hamprecht,
  Bengio and Courville}]{RahamanBaratin2019}
\bibinfo{author}{Rahaman, N.}, \bibinfo{author}{Baratin, A.},
  \bibinfo{author}{Arpit, D.}, \bibinfo{author}{Draxler, F.},
  \bibinfo{author}{Lin, M.}, \bibinfo{author}{Hamprecht, F.A.},
  \bibinfo{author}{Bengio, Y.}, \bibinfo{author}{Courville, A.C.},
  \bibinfo{year}{2019}.
\newblock \bibinfo{title}{On the spectral bias of neural networks}, in:
  \bibinfo{editor}{Chaudhuri, K.}, \bibinfo{editor}{Salakhutdinov, R.} (Eds.),
  \bibinfo{booktitle}{Proceedings of the 36th International Conference on
  Machine Learning, {ICML} 2019, 9-15 June 2019, Long Beach, California,
  {USA}}, \bibinfo{publisher}{{PMLR}}. pp. \bibinfo{pages}{5301--5310}.
\bibitem[{Raissi et~al.(2019)Raissi, Perdikaris and
  Karniadakis}]{RaissiPerdikaris2019}
\bibinfo{author}{Raissi, M.}, \bibinfo{author}{Perdikaris, P.},
  \bibinfo{author}{Karniadakis, G.}, \bibinfo{year}{2019}.
\newblock \bibinfo{title}{Physics-informed neural networks: A deep learning
  framework for solving forward and inverse problems involving nonlinear
  partial differential equations}.
\newblock \bibinfo{journal}{Journal of Computational Physics}
  \bibinfo{volume}{378}, \bibinfo{pages}{686--707}.
\bibitem[{Raissi et~al.(2020)Raissi, Yazdani and
  Karniadakis}]{RaissiYazdani2020}
\bibinfo{author}{Raissi, M.}, \bibinfo{author}{Yazdani, A.},
  \bibinfo{author}{Karniadakis, G.E.}, \bibinfo{year}{2020}.
\newblock \bibinfo{title}{Hidden fluid mechanics: Learning velocity and
  pressure fields from flow visualizations}.
\newblock \bibinfo{journal}{Science} \bibinfo{volume}{367},
  \bibinfo{pages}{1026--1030}.
\bibitem[{Rygiel et~al.(2023)Rygiel, P\l{}uszka, Zi\c{e}ba and
  Konopczy\'{n}ski}]{RygielPluszka2023}
\bibinfo{author}{Rygiel, P.}, \bibinfo{author}{P\l{}uszka, P.},
  \bibinfo{author}{Zi\c{e}ba, M.}, \bibinfo{author}{Konopczy\'{n}ski, T.},
  \bibinfo{year}{2023}.
\newblock \bibinfo{title}{{CenterlinePointNet++}: A new point cloud based
  architecture for coronary artery pressure drop and {vFFR} estimation}, in:
  \bibinfo{booktitle}{Medical Image Computing and Computer Assisted
  Intervention – MICCAI 2023: 26th International Conference, Vancouver, BC,
  Canada, October 8–12, 2023, Proceedings, Part VII},
  \bibinfo{publisher}{Springer-Verlag}, \bibinfo{address}{Berlin, Heidelberg}.
  p. \bibinfo{pages}{781–790}.
\bibitem[{Sharma et~al.(2012)Sharma, Itu, Zheng, Kamen, Bernhardt, Suciu and
  Comaniciu}]{SharmaItu2012}
\bibinfo{author}{Sharma, P.}, \bibinfo{author}{Itu, L.},
  \bibinfo{author}{Zheng, X.}, \bibinfo{author}{Kamen, A.},
  \bibinfo{author}{Bernhardt, D.}, \bibinfo{author}{Suciu, C.},
  \bibinfo{author}{Comaniciu, D.}, \bibinfo{year}{2012}.
\newblock \bibinfo{title}{A framework for personalization of coronary flow
  computations during rest and hyperemia}, in: \bibinfo{booktitle}{2012 Annual
  International Conference of the IEEE Engineering in Medicine and Biology
  Society}, \bibinfo{organization}{IEEE}. pp. \bibinfo{pages}{6665--6668}.
\bibitem[{Suk et~al.(2024a)Suk, Alblas, Hutten, Wiegman, Brune, van Ooij and
  Wolterink}]{SukAlblas2024}
\bibinfo{author}{Suk, J.}, \bibinfo{author}{Alblas, D.},
  \bibinfo{author}{Hutten, B.A.}, \bibinfo{author}{Wiegman, A.},
  \bibinfo{author}{Brune, C.}, \bibinfo{author}{van Ooij, P.},
  \bibinfo{author}{Wolterink, J.M.}, \bibinfo{year}{2024}a.
\newblock \bibinfo{title}{Physics-informed graph neural networks for flow field
  estimation in carotid arteries}, in: \bibinfo{booktitle}{n/a}, pp.
  \bibinfo{pages}{1--10}.
\bibitem[{Suk et~al.(2023)Suk, Brune and Wolterink}]{SukBrune2023}
\bibinfo{author}{Suk, J.}, \bibinfo{author}{Brune, C.},
  \bibinfo{author}{Wolterink, J.M.}, \bibinfo{year}{2023}.
\newblock \bibinfo{title}{{SE(3)} symmetry lets graph neural networks learn
  arterial velocity estimation from small datasets}, in:
  \bibinfo{editor}{Bernard, O.}, \bibinfo{editor}{Clarysse, P.},
  \bibinfo{editor}{Duchateau, N.}, \bibinfo{editor}{Ohayon, J.},
  \bibinfo{editor}{Viallon, M.} (Eds.), \bibinfo{booktitle}{Functional Imaging
  and Modeling of the Heart}, \bibinfo{publisher}{Springer Nature Switzerland},
  \bibinfo{address}{Cham}. pp. \bibinfo{pages}{445--454}.
\bibitem[{Suk et~al.(2024b)Suk, {de Haan}, Lippe, Brune and
  Wolterink}]{SukHaan2024a}
\bibinfo{author}{Suk, J.}, \bibinfo{author}{{de Haan}, P.},
  \bibinfo{author}{Lippe, P.}, \bibinfo{author}{Brune, C.},
  \bibinfo{author}{Wolterink, J.M.}, \bibinfo{year}{2024}b.
\newblock \bibinfo{title}{Mesh neural networks for {SE(3)}-equivariant
  hemodynamics estimation on the artery wall}.
\newblock \bibinfo{journal}{Computers in Biology and Medicine}
  \bibinfo{volume}{173}, \bibinfo{pages}{108328}.
\bibitem[{Suk et~al.(2024c)Suk, Haan, Imre and Wolterink}]{SukHaan2024b}
\bibinfo{author}{Suk, J.}, \bibinfo{author}{Haan, P.D.}, \bibinfo{author}{Imre,
  B.}, \bibinfo{author}{Wolterink, J.M.}, \bibinfo{year}{2024}c.
\newblock \bibinfo{title}{Geometric algebra transformers for large 3d meshes
  via cross-attention}, in: \bibinfo{booktitle}{ICML 2024 Workshop on
  Geometry-grounded Representation Learning and Generative Modeling}, pp.
  \bibinfo{pages}{0--6}.
\bibitem[{Suk et~al.(2024d)Suk, Imre and Wolterink}]{SukImre2024}
\bibinfo{author}{Suk, J.}, \bibinfo{author}{Imre, B.},
  \bibinfo{author}{Wolterink, J.M.}, \bibinfo{year}{2024}d.
\newblock \bibinfo{title}{Lab-gatr: geometric algebra transformers for large
  biomedical surface and volume meshes}.
\newblock \bibinfo{journal}{ArXiv} \bibinfo{volume}{abs/2403.07536}.
\bibitem[{Updegrove et~al.(2017)Updegrove, Wilson, Merkow, Lan, Marsden and
  Shadden}]{SimVascular2017}
\bibinfo{author}{Updegrove, A.}, \bibinfo{author}{Wilson, N.M.},
  \bibinfo{author}{Merkow, J.}, \bibinfo{author}{Lan, H.},
  \bibinfo{author}{Marsden, A.L.}, \bibinfo{author}{Shadden, S.C.},
  \bibinfo{year}{2017}.
\newblock \bibinfo{title}{{SimVascular: An Open Source Pipeline for
  Cardiovascular Simulation}}.
\newblock \bibinfo{journal}{Annals of Biomedical Engineering}
  \bibinfo{volume}{19}, \bibinfo{pages}{525--541}.
\bibitem[{Valen-Sendstad et~al.(2018)Valen-Sendstad, Bergersen, Shimogonya,
  Goubergrits, Bruening, Pallares, Cito, Piskin, Pekkan, Geers, Larrabide,
  Rapaka, Mihalef, Fu, Qiao, Jain, Roller, Mardal, Kamakoti, Spirka, Ashton,
  Revell, Aristokleous, Houston, Tsuji, Ishida, Menon, Browne, Broderick,
  Shojima, Koizumi, Barbour, Aliseda, Morales, Lefevre, Hodis, Al-Smadi, Tran,
  Marsden, Vaippummadhom, Einstein, Brown, Debus, Niizuma, Rashad, ichiro
  Sugiyama, Khan, Updegrove, Shadden, Cornelissen, Majoie, Berg, Saalfield,
  Kono and Steinman}]{ValenSendstadBergersen2018}
\bibinfo{author}{Valen-Sendstad, K.}, \bibinfo{author}{Bergersen, A.},
  \bibinfo{author}{Shimogonya, Y.}, \bibinfo{author}{Goubergrits, L.},
  \bibinfo{author}{Bruening, J.}, \bibinfo{author}{Pallares, J.},
  \bibinfo{author}{Cito, S.}, \bibinfo{author}{Piskin, S.},
  \bibinfo{author}{Pekkan, K.}, \bibinfo{author}{Geers, A.},
  \bibinfo{author}{Larrabide, I.}, \bibinfo{author}{Rapaka, S.},
  \bibinfo{author}{Mihalef, V.}, \bibinfo{author}{Fu, W.},
  \bibinfo{author}{Qiao, A.}, \bibinfo{author}{Jain, K.},
  \bibinfo{author}{Roller, S.}, \bibinfo{author}{Mardal, K.A.},
  \bibinfo{author}{Kamakoti, R.}, \bibinfo{author}{Spirka, T.},
  \bibinfo{author}{Ashton, N.}, \bibinfo{author}{Revell, A.},
  \bibinfo{author}{Aristokleous, N.}, \bibinfo{author}{Houston, J.},
  \bibinfo{author}{Tsuji, M.}, \bibinfo{author}{Ishida, F.},
  \bibinfo{author}{Menon, P.}, \bibinfo{author}{Browne, L.},
  \bibinfo{author}{Broderick, S.}, \bibinfo{author}{Shojima, M.},
  \bibinfo{author}{Koizumi, S.}, \bibinfo{author}{Barbour, M.},
  \bibinfo{author}{Aliseda, A.}, \bibinfo{author}{Morales, H.},
  \bibinfo{author}{Lefevre, T.}, \bibinfo{author}{Hodis, S.},
  \bibinfo{author}{Al-Smadi, Y.}, \bibinfo{author}{Tran, J.},
  \bibinfo{author}{Marsden, A.}, \bibinfo{author}{Vaippummadhom, S.},
  \bibinfo{author}{Einstein, G.}, \bibinfo{author}{Brown, A.},
  \bibinfo{author}{Debus, K.}, \bibinfo{author}{Niizuma, K.},
  \bibinfo{author}{Rashad, S.}, \bibinfo{author}{ichiro Sugiyama, S.},
  \bibinfo{author}{Khan, M.}, \bibinfo{author}{Updegrove, A.},
  \bibinfo{author}{Shadden, S.}, \bibinfo{author}{Cornelissen, B.},
  \bibinfo{author}{Majoie, C.}, \bibinfo{author}{Berg, P.},
  \bibinfo{author}{Saalfield, S.}, \bibinfo{author}{Kono, K.},
  \bibinfo{author}{Steinman, D.}, \bibinfo{year}{2018}.
\newblock \bibinfo{title}{Real-world variability in the prediction of
  intracranial aneurysm wall shear stress: The 2015 international aneurysm cfd
  challenge}.
\newblock \bibinfo{journal}{Cardiovascular engineering and technology}
  \bibinfo{volume}{9}, \bibinfo{pages}{544--564}.
\bibitem[{Vaswani et~al.(2017)Vaswani, Shazeer, Parmar, Uszkoreit, Jones,
  Gomez, Kaiser and Polosukhin}]{VaswaniShazeer2017}
\bibinfo{author}{Vaswani, A.}, \bibinfo{author}{Shazeer, N.},
  \bibinfo{author}{Parmar, N.}, \bibinfo{author}{Uszkoreit, J.},
  \bibinfo{author}{Jones, L.}, \bibinfo{author}{Gomez, A.N.},
  \bibinfo{author}{Kaiser, L.}, \bibinfo{author}{Polosukhin, I.},
  \bibinfo{year}{2017}.
\newblock \bibinfo{title}{Attention is all you need}, in:
  \bibinfo{editor}{Guyon, I.}, \bibinfo{editor}{von Luxburg, U.},
  \bibinfo{editor}{Bengio, S.}, \bibinfo{editor}{Wallach, H.M.},
  \bibinfo{editor}{Fergus, R.}, \bibinfo{editor}{Vishwanathan, S.V.N.},
  \bibinfo{editor}{Garnett, R.} (Eds.), \bibinfo{booktitle}{Advances in Neural
  Information Processing Systems 30: Annual Conference on Neural Information
  Processing Systems 2017, December 4-9, 2017, Long Beach, CA, {USA}}, pp.
  \bibinfo{pages}{5998--6008}.
\bibitem[{Wang et~al.(2023)Wang, Wu, Li, Zhang, Xiao, Li, Qiao, Jin and
  Liu}]{WangWu2023}
\bibinfo{author}{Wang, S.}, \bibinfo{author}{Wu, D.}, \bibinfo{author}{Li, G.},
  \bibinfo{author}{Zhang, Z.}, \bibinfo{author}{Xiao, W.}, \bibinfo{author}{Li,
  R.}, \bibinfo{author}{Qiao, A.}, \bibinfo{author}{Jin, L.},
  \bibinfo{author}{Liu, H.}, \bibinfo{year}{2023}.
\newblock \bibinfo{title}{Deep learning-based hemodynamic prediction of carotid
  artery stenosis before and after surgical treatments}.
\newblock \bibinfo{journal}{Frontiers in Physiology} \bibinfo{volume}{13},
  \bibinfo{pages}{1094743}.
\bibitem[{Wessels et~al.(2025)Wessels, Knigge, Valperga, Papa, Vadgama, Gavves
  and Bekkers}]{WesselsKnigge2025}
\bibinfo{author}{Wessels, D.}, \bibinfo{author}{Knigge, D.M.},
  \bibinfo{author}{Valperga, R.}, \bibinfo{author}{Papa, S.},
  \bibinfo{author}{Vadgama, S.}, \bibinfo{author}{Gavves, E.},
  \bibinfo{author}{Bekkers, E.J.}, \bibinfo{year}{2025}.
\newblock \bibinfo{title}{Grounding continuous representations in geometry:
  Equivariant neural fields}, in: \bibinfo{booktitle}{The Thirteenth
  International Conference on Learning Representations}, pp.
  \bibinfo{pages}{0--21}.
\bibitem[{Wu et~al.(2024)Wu, Wu, Theodorou, Liang, Mack, Glass, Sun and
  Zou}]{WuWu2024}
\bibinfo{author}{Wu, K.}, \bibinfo{author}{Wu, E.}, \bibinfo{author}{Theodorou,
  B.}, \bibinfo{author}{Liang, W.}, \bibinfo{author}{Mack, C.},
  \bibinfo{author}{Glass, L.}, \bibinfo{author}{Sun, J.}, \bibinfo{author}{Zou,
  J.}, \bibinfo{year}{2024}.
\newblock \bibinfo{title}{Characterizing the clinical adoption of medical {AI}
  devices through {U.S.} insurance claims}.
\newblock \bibinfo{journal}{NEJM AI} \bibinfo{volume}{1},
  \bibinfo{pages}{AIoa2300030}.
\bibitem[{Xie et~al.(2022)Xie, Takikawa, Saito, Litany, Yan, Khan, Tombari,
  Tompkin, sitzmann and Sridhar}]{XieTakikawa2022}
\bibinfo{author}{Xie, Y.}, \bibinfo{author}{Takikawa, T.},
  \bibinfo{author}{Saito, S.}, \bibinfo{author}{Litany, O.},
  \bibinfo{author}{Yan, S.}, \bibinfo{author}{Khan, N.},
  \bibinfo{author}{Tombari, F.}, \bibinfo{author}{Tompkin, J.},
  \bibinfo{author}{sitzmann, V.}, \bibinfo{author}{Sridhar, S.},
  \bibinfo{year}{2022}.
\newblock \bibinfo{title}{Neural fields in visual computing and beyond}.
\newblock \bibinfo{journal}{Computer Graphics Forum} \bibinfo{volume}{41},
  \bibinfo{pages}{641--676}.
\bibitem[{Zhang et~al.(2023)Zhang, Mao, Che, Kang, Luo, Qiao, Liu, Anzai, Ohta,
  Guo and Li}]{ZhangMao2023}
\bibinfo{author}{Zhang, X.}, \bibinfo{author}{Mao, B.}, \bibinfo{author}{Che,
  Y.}, \bibinfo{author}{Kang, J.}, \bibinfo{author}{Luo, M.},
  \bibinfo{author}{Qiao, A.}, \bibinfo{author}{Liu, Y.},
  \bibinfo{author}{Anzai, H.}, \bibinfo{author}{Ohta, M.},
  \bibinfo{author}{Guo, Y.}, \bibinfo{author}{Li, G.}, \bibinfo{year}{2023}.
\newblock \bibinfo{title}{Physics-informed neural networks (pinns) for 4d
  hemodynamics prediction: An investigation of optimal framework based on
  vascular morphology}.
\newblock \bibinfo{journal}{Computers in Biology and Medicine}
  \bibinfo{volume}{164}, \bibinfo{pages}{107287}.
\bibitem[{Zingaro et~al.(2023)Zingaro, Vergara, Dede, Regazzoni and
  Quarteroni}]{ZingaroVergara2023}
\bibinfo{author}{Zingaro, A.}, \bibinfo{author}{Vergara, C.},
  \bibinfo{author}{Dede, L.}, \bibinfo{author}{Regazzoni, F.},
  \bibinfo{author}{Quarteroni, A.}, \bibinfo{year}{2023}.
\newblock \bibinfo{title}{A comprehensive mathematical model for cardiac
  perfusion}.
\newblock \bibinfo{journal}{Scientific Reports} \bibinfo{volume}{13}.

\end{thebibliography}

\clearpage

\revision{}{\section{Preliminary results on wall shear stress}}
\begin{figure*}
	\centering
	\includegraphics{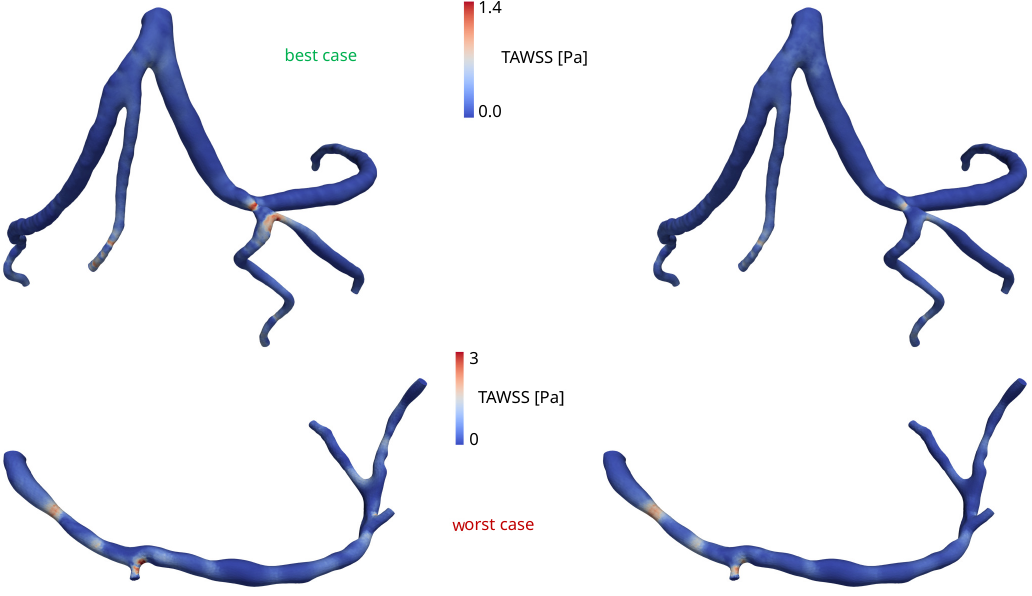}
	\caption{\revision{}{\textbf{Preliminary results} on time-averaged wall shear stress (TAWSS) extraction from the estimated velocity field at different levels of accuracy (w.r.t. MAE associated with TAWSS) in test subjects. The backend model was LaB-GATr.}}\label{fig:preliminary}
\end{figure*}

\begin{figure}
	\centering
	\includegraphics{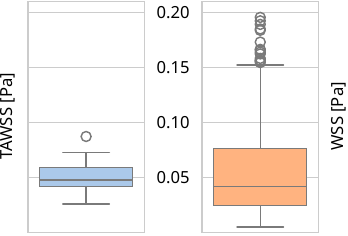}
	\caption{\revision{}{\textbf{Distribution of MAE} $\downarrow$ of time-averaged wall shear stress (TAWSS) (over subjects) and wall shear stress (WSS) (over time and subjects) between the machine learning model compared to CFD as boxplots for the test split. The backend model was LaB-GATr.}}\label{fig:wss}
\end{figure}

\revision{}{In Figure~\ref{fig:preliminary} we show preliminary results on time-averaged wall shear stress (TAWSS) extraction from the estimated velocity fields for two levels of accuracy. We compute wall shear stress (WSS) via $\mathrm{L}^2$ projection based on the tetrahedral mesh from CFD simulation. We observe that high TAWSS is captured in terms of artery segment but not magnitude. We attribute this to the fact that wall shear stress is a first-order differential quantity for which we do not yet account for via velocity field regularisation during training. In Figure~\ref{fig:wss} we provide comparison of the distribution of MAE for TAWSS and WSS across the test split.}



\end{document}